\documentclass[12pt]{article}
\usepackage{amsmath}
\usepackage{amsthm}
\usepackage{amssymb}
\usepackage{graphicx}
\usepackage{enumerate}
\usepackage{natbib}
\usepackage{url} 
\usepackage{bbm}
\newtheorem{remark}{Remark}
\newtheorem{theorem}{Theorem}[section]

\begin{document}

\def\spacingset#1{\renewcommand{\baselinestretch}%
{#1}\small\normalsize} \spacingset{1}


\title{\bf Occam Factor for Random Graphs: Erd\"os-R\'enyi, Independent Edge, and Rank-1 Stochastic Blockmodel}
\author{
Tianyu Wang\thanks{This work is partially supported by NIH MH-19-147, and by Microsoft Research.}\\
Zachary M. Pisano \\
Carey E. Priebe \\
Department of Applied Mathematics and Statistics, Johns Hopkins University}
\maketitle

\bigskip
\begin{abstract}
We investigate the evidence/flexibility (i.e., ``Occam'') paradigm and demonstrate the theoretical and empirical consistency of Bayesian evidence for the task of determining an appropriate generative model for network data. This model selection framework involves determining a collection of candidate models, equipping each of these models' parameters with prior distributions derived via the encompassing priors method, and computing or approximating each models' evidence. We demonstrate how such a criterion may be used to select the most suitable model among the Erd\"{o}s-R\'{e}nyi (ER) model, independent edge (IE) model, and rank-1 stochastic blockmodel (SBM). The Erd\"{o}s-R\'{e}nyi may be considered as being linearly nested within IE, a fact which permits exponential family results. The rank-1 SBM is not so ideal, so we propose a numerical method to approximate its evidence. We apply this paradigm to brain connectome data. Future work necessitates deriving and equipping additional candidate random graph models with appropriate priors so they may be included in the paradigm.
\end{abstract}

\noindent%
\vfill

\newpage
\spacingset{1.9} 

\section{Introduction}

Network data have become increasingly emphasized in the digital age, thus necessitating the development of statistical methods by which they may be analyzed. The modeling of networks via random graphs parameterized by edge probability matrices has proven fruitful in this regard. One particular challenge faced by researchers is that such data in their most general form possess high dimension; simplifications of the general case have been proposed, in particular the random dot product graph (RDPG), stochastic blockmodel (SBM), and Erd\"{o}s-R\'{e}nyi models \citep{JMLR}. Should one determine if a given network may be more suitably modeled by one of these more specific settings, one may then advantageously utilize analytical and inferential methods appropriate to that selection. 

The underlying probabilistic structure of random graph models has posed technical difficulties for the development of such methods. As outlined by \cite{JMLR}, it is generally more helpful to consider low-dimensional spectral embeddings of an observed graph's adjacency matrix. This approach may be accompanied by its own confounding issues, namely that the rows of these embeddings are subject to identifiability only up to orthogonal transformation, and also are subject to curved-Gaussian asymptotic results. \cite{Pisano-2022} took advantage of this curved-Gaussian limiting structure in SBMs via an expectation-solution algorithm for the spectral clustering of graph vertices which improved upon a similar procedure conducted via the expectation-maximization algorithm.

However, such inferential tasks are downstream from that of model selection, which for these settings has historically involved the use of BIC and Bayesian evidence pertaining to these Gaussian asymptotics \citep{yang-21,passino-2020}. The purpose of this article and subsequent work (highlighted in the Discussion below) is develop to a means of model selection which directly invokes the structure of random graph models without necessarily appealing to results concerning asymptotic Gaussianity of low-rank representations of network data. We propose examining this problem with the Occam paradigm introduced by \cite{Priebe-2019} and expanded upon in Chapter 3 of \cite{Pisano-diss}. This paradigm is founded upon the principle that the frequentist framework of model selection involves --- either implicitly or explicitly --- some form of regularization of the various models being considered; while the exclusion of certain regularization terms from a model selection criterion may not affect that criterion's consistency in the limit --- see, e.g., BIC's consistency as originally established by \cite{BIC} --- in the presence of only moderate amounts of data this may not be the case \citep{Gelf_Dey-1994}. We frequently refer to the Occam paradigm as ``the evidence/flexibility paradigm'' throughout the article; the ``Occamization'' of a model selection problem consists of first deciding upon a collection of candidate models, equipping the simplest model with an appropriate and defensible prior, equipping the remaining models by means of the encompassing priors method due to \cite{Klugkist-2007}, deriving or approximating the posterior distribution for each model's parameter(s), and then computing or approximating each model's evidence.

\begin{figure}[h!]
    \centering
    \includegraphics[width=.8\linewidth]{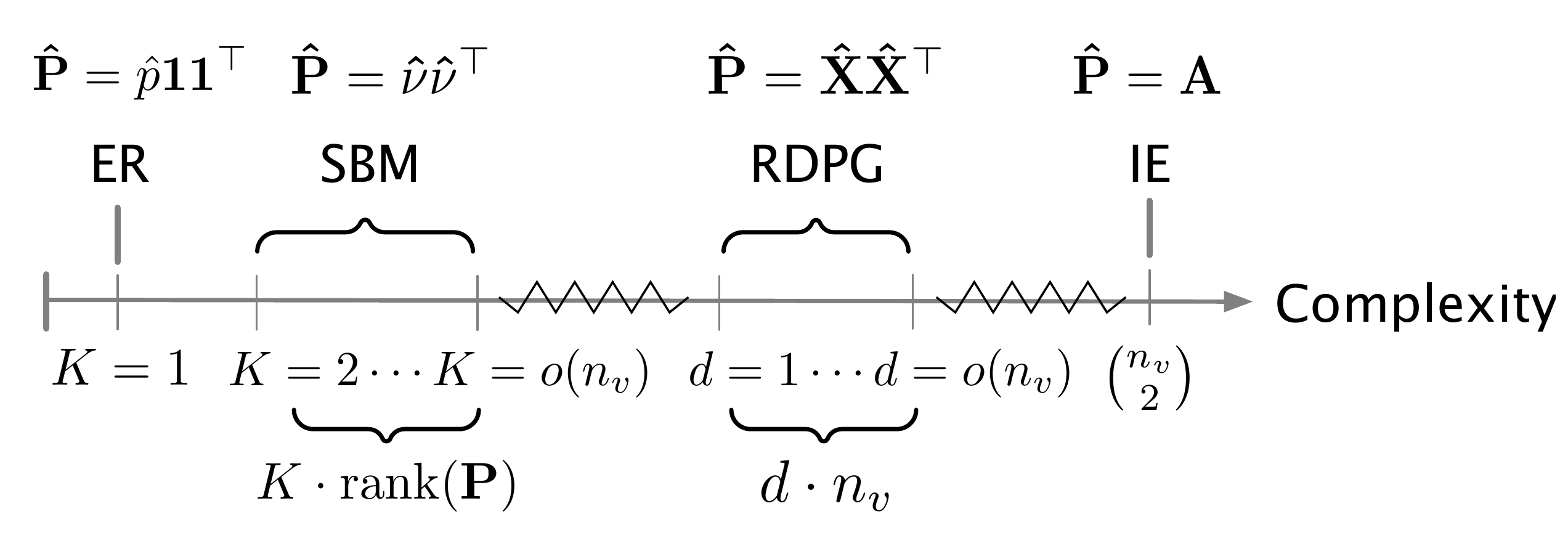}
    \caption{Model complexity for various random graph models. Here, $K$ is the number of blocks in an SBM; note that when an SBM with $K=1$ is an ER model. Here also $d$ is the true rank of the edge probability matrix \textbf{P}.}
    \label{fig:fit-complexity}
\end{figure}

For certain special cases of the model selection problem --- e.g., choosing among multiple full-rank exponential family models, which may be ordered according to a linear nesting --- all terms in the Occam paradigm may be computed exactly. Moreover, for such settings once can prescribe a means of matching hyperparameters across the various models \textit{\`a la} Theorem \ref{third_theorem}. We briefly review the necessary results in Section \ref{sec:background}. These theoretical considerations precisely apply to the problem of selecting between the Erd\"{o}s-R\'{e}nyi (ER) and independent-edge (IE) graph models, which respectively represent the simplest and most complex graph models one might consider (Section \ref{sec:graphs_er_vs_ie}). Other models, such as SBMs and RDPGs, reside within the ``complexity gap'' between these two extremes shown in Figure \ref{fig:fit-complexity}. We begin the task of ``Occamizing'' this gap in Section \ref{sec:sbm_known} and \ref{sec:sbm_unknown}, which concerns the inclusion of rank-1 SBMs; Bayesian evidence in this case cannot be computed exactly, so we turn to numerical methods. In Section \ref{sec:simulation} we illustrate our evidence-based criterion with simulation. Then we present an application of this paradigm to brain connectome data in Section \ref{sec:application} and conclude with a discussion in Section \ref{sec:discussion}.

\section{Evidence and Flexibility}\label{sec:background}

The relationship between evidence and flexibility is succinctly derived from the usual Bayes rule. For a likelihood model $L(\boldsymbol{\psi})$ with prior distribution $\rho(\boldsymbol{\psi})$, the posterior distribution of $\boldsymbol{\psi}$ given the data $\boldsymbol{x}$ is computed as
\begin{equation}\label{Bayes_rule}
    \rho(\boldsymbol{\psi} | \boldsymbol{x}) = \frac{L(\boldsymbol{\psi})\rho(\boldsymbol{\psi})}{\int_{\boldsymbol\Psi}L(\boldsymbol{\psi})\rho(\boldsymbol{\psi}) d\boldsymbol{\psi}}.
\end{equation}
Let $E$ denote the \textit{evidence}, i.e., the integral in the denominator on the right hand side. Upon rearranging and taking logs we obtain
\begin{equation}\label{log_evi}
    \log E = \log L(\boldsymbol{\psi}) - \log \frac{\rho(\boldsymbol{\psi} | \boldsymbol{x})}{\rho(\boldsymbol{\psi})},
\end{equation}
yielding the log-evidence as the log-likelihood minus a penalty expression, dubbed \textit{flexibility} by \cite{Priebe-2019}, and thus exactifying an approximation found in \cite{MacKay-1992}. We denote the flexibility here and throughout as $\mathcal{F}(\boldsymbol\psi, \boldsymbol x)$.

The expression (\ref{log_evi}) holds for all $\boldsymbol\psi \in \boldsymbol{\Psi}$, hence $\log E$ as a criterion may be interpreted as evaluating the entirety of a model given the observed data; in practice one need not even estimate the parameter --- if both the likelihood and flexibility possess known forms, one may simply plug in any value which greatly simplifies the calculation. We contrast this with BIC, which evaluates fit at the ML or MAP estimate of $\boldsymbol\psi$, depending on whether we equip our model with a regularizer/prior. Nonetheless we can compare the BIC penalty with flexibility by evaluating the latter at the MAP estimate
\begin{equation*}
    \hat{\boldsymbol{\psi}} := \text{argmax}_{\boldsymbol\psi \in \boldsymbol\Psi}(\log L(\boldsymbol\psi) + \log\rho(\boldsymbol\psi)).
\end{equation*}
Frequentist justification of flexibility as a likelihood penalty and evidence as a model selection criterion arises in the context of the regularized maximum likelihood procedure
\begin{equation*}
    \check{\boldsymbol{\psi}} := \text{argmax}_{\boldsymbol\psi \in \boldsymbol\Psi}(\log L(\boldsymbol\psi) - R(\boldsymbol\psi))
\end{equation*}
where $R$ is a regularizer functionally equivalent to a prior
\begin{equation}\label{prior_reg}
    \rho_R(\boldsymbol\psi) \propto \exp{\left(-R(\boldsymbol\psi)\right)}.
\end{equation}

\subsection{Evidence and Flexibility for a Canonical Exponential Family}\label{2.1}

Let us consider the general form flexibility exhibits for an i.i.d. sample from a canonical $k$-rank exponential family equipped with a conjugate prior. Denote the density of the $i$th observation as $f(x_i | \boldsymbol\psi) = h(x_i)\exp{\left(\langle\boldsymbol\theta, \boldsymbol T(x_i)\rangle - A(\boldsymbol\theta)\right)}$, with base measure $h(\cdot)$, canonical parameter $\boldsymbol\theta := \boldsymbol\theta(\boldsymbol\psi)$, sufficient statistic $\boldsymbol T(\cdot)$, and log-partition function $A(\boldsymbol\theta)$. We have immediately that $L(\boldsymbol\psi) = \big(\prod_{i=1}^n h(x_i)\big) \exp{\left(\langle\boldsymbol\theta,  \sum_{i=1}^n\boldsymbol T(x_i)\rangle - nA(\boldsymbol\theta)\right)}$. Per \cite{Diaconis-1979}, the natural conjugate prior for $\boldsymbol\theta$ has the form 
\begin{equation}\label{conjugate_prior}
    \rho(\boldsymbol\theta) = H(\boldsymbol\tau, m)\exp{\left(\langle\boldsymbol\tau, \boldsymbol\theta\rangle - mA(\boldsymbol\theta)\right)},
\end{equation}
where the normalizing contant $H$ depends upon the hyperparameters $\boldsymbol\tau$ and $m$ which respectively act as a prior estimate for the sufficient statistic and the prior sample size. This prior corresponds to the ``conjugate regularizer'' $R(\boldsymbol{\psi}) = mA(\boldsymbol\theta) - \langle\boldsymbol\tau, \boldsymbol\theta\rangle$; by appealing to the properties of the log-partition one intuits that the first term penalizes parameter values near the infinitely-valued boundary of the canonical parameter space $\{\boldsymbol\theta : A(\boldsymbol\theta)<\infty\}$ by a degree equal to $m$, and the second term deemphasizes values angularly different from $\boldsymbol\tau$. To see that this prior is indeed conjugate, observe that computing the posterior in the usual way reveals that the hyperparameters may be updated by the simple rule
\begin{align*}
    \boldsymbol\tau &\to \sum_{i=1}^n \boldsymbol T(x_i) + \boldsymbol\tau\\
    m &\to n+m.
\end{align*}

Under regularity conditions analogous to those yielding Theorem 2.3.1 in \cite{Bickel-Doksum-2015} the MAP estimate of the canonical parameter exists uniquely and is equal to $\boldsymbol{\hat\theta} = \dot A^{-1}\big((\sum_{i=1}^n \boldsymbol T(x_i) +\boldsymbol\tau)/(n+m) \big)$, and the corresponding flexibility is
\begin{equation}\label{flex_exp_fam}
    \mathcal F(\boldsymbol{\hat\theta}, \boldsymbol x) = \log \frac{H\big(\sum_{i=1}^n \boldsymbol T(x_i) +\boldsymbol\tau, n+m\big)}{H(\boldsymbol\tau, m)} + \langle\boldsymbol{\hat\theta}, \sum_{i=1}^n\boldsymbol T(x_i)\rangle - nA(\boldsymbol{\hat\theta}).
\end{equation}
From (\ref{log_evi}) and (\ref{flex_exp_fam}) one immediately deduces that the evidence in the exponential family case may be written
\begin{equation}\label{evi}
    E = \frac{\big(\prod_{i=1}^n h(x_i)\big)H(\boldsymbol\tau, m)}{H\big(\sum_{i=1}^n \boldsymbol T(x_i) +\boldsymbol\tau, n+m\big)}
\end{equation}
The following result relates the asymptotic behavior of flexibility with that of the BIC penalty $(k/2)\log n$.
\begin{theorem}\label{Op1_theorem}
Suppose the observations $x_1,\dots,x_n$ are generated from the aforementioned exponential family equipped with conjugate prior $\rho(\cdot)$. Define $t$ such that $\sum_{i=1}^n\boldsymbol T(x_i)/n \overset{p}\to t$, and assume that $\boldsymbol\Theta$ is open, $m>0$, and $\boldsymbol\tau/m$ is in the convex support of $\boldsymbol T(X)$. By the continuous mapping theorem, there exists $\boldsymbol\theta_0$ such that the sequence of MAP estimates $\boldsymbol{\hat\theta}_n \overset{p}\to \boldsymbol\theta_0$ as $n\to\infty$, and we have that
\begin{equation*}
    \mathcal F(\boldsymbol{\hat\theta}_n, \boldsymbol{x}) - \frac{k}{2}\log n \overset{p}\to -\log \rho(\boldsymbol\theta_0) + \frac12 \log\bigg|\frac{\overset{\boldsymbol{\cdot\cdot}}A(\boldsymbol\theta_0)}{2\pi}\bigg|
\end{equation*}
and consequently
\begin{equation*}
    \log E - BIC(\boldsymbol{\hat\theta}_n) \overset{p}\to -\log \rho(\boldsymbol\theta_0) + \frac12 \log\bigg|\frac{\overset{\boldsymbol{\cdot\cdot}}A(\boldsymbol\theta_0)}{2\pi}\bigg|
\end{equation*}
as well.
\end{theorem}
The proof may be found in the Appendix A.1, and largely depends upon a Laplace approximation to the posterior distribution around the MAP estimate. This result yields the constant which the BIC penalty must dominate to validate the heuristic ``log-evidence $\approx$ BIC''. Given the slow growth of the logarithmic function (\citealp*{Gowers-2002}, p.117), the number of observations needed to permit this approximation might be truly large in some settings.
Theorem \ref{Op1_theorem} hinges on defining BIC as 
\begin{equation}\label{BIC}
    \log L(\boldsymbol{\hat\theta}) - \frac k{2}\log n,
\end{equation}
which is incorrect when $\rho$ is not informative; in such case the MAP estimate of $\boldsymbol\theta$ is also the ML estimate. The correct expression for BIC in the presence of a prior distribution is
\begin{equation}\label{prior_corrected_BIC}
    \text{BIC} = \log L(\boldsymbol{\hat\theta}) + \log\rho(\boldsymbol{\hat\theta}) - \frac{k}{2}\log n,
\end{equation}
hereafter referred to as the \textit{prior-corrected BIC}. An immediate consequence of Theorem \ref{Op1_theorem} is that the difference between flexibility and the correct BIC penalty reduces to $\log\rho(\boldsymbol{\hat\theta}|\boldsymbol x) - (k/2)\log n$, which converges in probability to $0.5\log|\overset{\boldsymbol{\cdot\cdot}}A(\boldsymbol\theta_0)/(2\pi)|$. Again, the sample size necessary for the BIC penalty to dominate this term might be excessively large; Table 3.4 in \cite{Pisano-diss} demonstrates this gap for an example with $n=203$. To account for this limiting constant one might instead consider the penalty $\kappa(\boldsymbol{\hat\theta},\boldsymbol x) = (k/2)\log n -\log\rho(\boldsymbol{\hat\theta}) - 0.5\log\big|\overset{\boldsymbol{\cdot\cdot}}A(\boldsymbol{\hat\theta})(2\pi)\big|$, which yields the Kashyap information criterion \citep{Kashyap-1982} when subtracted from the log-likelihood. One immediately notes that $\mathcal F(\boldsymbol{\hat\theta}, \boldsymbol{x}) - \kappa(\boldsymbol{\hat\theta},\boldsymbol x) \overset{p}\to 0$.

\begin{remark}
Analogous results in this section may yet be obtained when we equip our likelihood setting with the flat and improper prior $\rho(\boldsymbol\theta) \propto 1$ on $\boldsymbol\Theta$ (in a Bayesian sense), or with no regularizer at all (in a frequentist sense). In this case estimation of and inference about $\boldsymbol\theta$ is performed via maximum likelihood. The evidence and flexibility may then be respectively computed as
\begin{equation*}
    E = \frac{\prod_{i=1}^n h(x_i)}{H(\sum_{i=1}^n \boldsymbol T(x_i), n)},
\end{equation*}
and
\begin{equation*}
    \mathcal F(\boldsymbol{\hat\theta}, \boldsymbol x) = \log H(\sum_{i=1}^n \boldsymbol T(x_i), n) + \langle\boldsymbol{\hat\theta}, \sum_{i=1}^n\boldsymbol T(x_i)\rangle - nA(\boldsymbol{\hat\theta}).
\end{equation*}
\end{remark}

\subsection{Conjugate Priors for Linearly Nested Submodels}\label{2.2}

We also define a linearly \textit{nested} submodel ($N$) of rank $\ell<k$ with density
\begin{equation*}
    g(x_i|\boldsymbol\eta) = h(x_i)\exp\left(\langle\textbf{M}^\top\boldsymbol\eta,\boldsymbol T(x_i)\rangle-A(\textbf{M}^\top\boldsymbol\eta)\right)
\end{equation*}
where $h, \boldsymbol T,$ and $A$ are as in the \textit{full} model ($F$) discussed in Section \ref{2.1}, $\textbf{M}\in \mathbbm R^{\ell\times k}$ is rank-$\ell$, and $\boldsymbol\eta$ takes values in $\mathcal E := \{\boldsymbol\eta\in\mathbbm R^\ell : |A(\textbf{M}^\top\boldsymbol\eta)|<\infty\}$, the natural canonical parameter space for the nested model. Note that the image $\textbf{M}^\top(\mathcal E)$ is a linear subset of the full model's natural parameter space $\boldsymbol\Theta$. Under such conditions the model $N$ is fully rank-$\ell$, with natural sufficient statistic $\textbf{M}\boldsymbol T(X)$, base measure $h$, and log-partition $B(\boldsymbol\eta) = A(\textbf{M}^\top\boldsymbol\eta)$; we have further that $\dot B(\boldsymbol\eta) = \textbf{M}\dot A(\textbf{M}^\top\boldsymbol\eta)$ is one-to-one on $\mathcal E$ via a well-known result concerning the properties of exponential families (e.g., \cite{Bickel-Doksum-2015}, Theorem 1.6.4, and problem 1.6.17). We can likewise define a conjugate prior for $\boldsymbol\eta$
\begin{equation*}
    \rho_N(\boldsymbol\eta) = G(\boldsymbol\upsilon, w)\exp\left(\langle\boldsymbol\eta,\boldsymbol\upsilon\rangle - wB(\boldsymbol\eta)\right)
\end{equation*}
in which the likelihood updates the hyperparameters via the rule
\begin{align*}
    \boldsymbol\upsilon &\to \textbf{M}\sum_{i=1}^n T(x_i) + \boldsymbol\upsilon\\
    w &\to n+w.
\end{align*}

In the presence of i.i.d. observations $x_1,\dots,x_n$ and conditions analogous to those leading to the unique existence of the MAP estimate for model $F$ --- namely that $\mathcal E$ is open and $(\textbf{M}\sum_{i=1}^n\boldsymbol T(x_i)+\boldsymbol\upsilon))/(n+w)$ lies in the interior of the convex support of the natural sufficient statistic --- the MAP estimate for $N$ exists uniquely as
\begin{equation*}
    \boldsymbol{\hat\eta}_n = \dot B^{-1}(\frac{\textbf{M}\sum_{i=1}^n\boldsymbol T(x_i)+\boldsymbol\upsilon}{n+w}).
\end{equation*}
The setting as we have described it thus far also leads us to conclude that $\dot B^{-1}$ is continuous on the interior of the convex support of $\textbf{M}\boldsymbol T(X)$, so even when model $F$ is true (i.e., the true parameter value $\boldsymbol\theta_0\in\boldsymbol\Theta \setminus \textbf{M}^\top(\mathcal E)$) the sequence of MAP estimators for model $N$ converges in probability to the value $\boldsymbol{\tilde\eta}_0 = \dot B^{-1}(\textbf{M}\mathbbm E_{\boldsymbol\theta_0}\lbrack \boldsymbol T(X)\rbrack) = \dot B^{-1}(\textbf{M}\dot A(\boldsymbol\theta_0)$. When $N$ is true (i.e., there exists $\boldsymbol\eta_0\in \mathcal E$ such that $\boldsymbol\theta_0 = \textbf{M}^\top \boldsymbol\eta_0$) we see immediately $\boldsymbol{\tilde\eta_0} = \dot B^{-1}(\textbf{M}\dot A(\textbf{M}^\top\boldsymbol\eta_0) = \dot B^{-1}(\dot B(\boldsymbol\eta_0)) = \boldsymbol\eta_0$.

One question that arose in the design of the simulations in Section 4 was: Assuming that we know the hyperparameters for the true model's prior, how should we specify the hyperparameters for the other priors? Certainly, such a choice ought not be arbitrary, since no matter the sample size one can hyperparameterize the incorrect priors in such a way that evidence chooses an incorrect model. Such a concern brings to mind \cite{edwards-84} objection to the Bayesian framework: ``It is sometimes said, in defence of the Bayesian concept, that the choice of prior distribution is unimportant in practice because it hardly influences the posterior distribution at all when there are moderate amounts of data. The less said about this 'defence' the better.''

To address this issue we recall the original interpretation of $m$ and $\boldsymbol\tau$ as the prior sample size and sufficient statistic, respectively. Therefore it seems reasonable to hyperparameterize the other priors in such a way that the information content of $m$ and $\boldsymbol\tau$ are preserved. For a linearly nested submodel characterized by the linear transformation $\textbf{M}$, it seems appropriate to simply take $w=m$ and $\boldsymbol\upsilon = \textbf{M}\boldsymbol\tau$; the prior for the nested model with this hyperparameterization possesses the interpretation of being ``closest'' to the true prior for the full model. Likewise, we can project the hyperparameterization $(\boldsymbol\upsilon, w)$ of a true nested model up to the full model by taking $m=w$ and $\boldsymbol\tau' =\textbf{M}^\top(\textbf{MM}^\top)^{-1}\boldsymbol\upsilon$; this yields the hyperparameterization for the full prior which is ``closest'' to the nested prior. We render this more explicit with the following information-theoretic result, and we show the proof in Appendix A.2.

\begin{theorem}\label{third_theorem}
Suppose $\rho_F$ and $\rho_N$ are as discussed above, and that both $\boldsymbol\tau$ and $m$ are known. Let $\mathbbm E_N$ denote expecation over $\rho_N$. We have that
\begin{equation*}
    \underset{\underset{w}{\boldsymbol\upsilon}}{\min} -\mathbbm E_{N}\bigg\lbrack\log\frac{\rho_F(\textbf{M}^\top\boldsymbol\eta)}{\rho_N(\boldsymbol\eta)}\bigg\rbrack = \log\frac{G(\textbf{M}\boldsymbol\tau, m)}{H(\boldsymbol\tau, m)}
\end{equation*}
is achieved by taking $\boldsymbol\upsilon= \textbf{M}\boldsymbol\tau$ and $w=m$.

Likewise, suppose instead that $\boldsymbol\upsilon$ and $w$ are known. Let $\textbf{M}^+$ denote the (right) Moore-Penrose pseudo-inverse of $\textbf{M}$, and define $\mathcal T(\boldsymbol\upsilon) = \{\boldsymbol\tau : \textbf{M}\boldsymbol\tau =\boldsymbol\upsilon\}$. We have that
\begin{equation*}
    \underset{\underset{m=w}{\boldsymbol\tau\in\mathcal T(\boldsymbol\upsilon)}}{\min} - \mathbbm E_{N}\bigg\lbrack\log\frac{\rho_F(\textbf{M}^\top\boldsymbol\eta)}{\rho_N(\boldsymbol\eta)}\bigg\rbrack = \log\frac{G(\boldsymbol\upsilon, w)}{H(\textbf{M}^+\boldsymbol\upsilon, w)}
\end{equation*}
is achieved by by taking $\boldsymbol\tau = \textbf{M}^+\boldsymbol\upsilon$.
\end{theorem}

That is to say that, the closest prior for the nested model to a given prior for the full model (in an information-theoretic sense) is that with hyperparameters $\boldsymbol\upsilon = \textbf{M}\boldsymbol\tau$ and $w=m$; and the closest full prior to a given nested prior among all such full priors for which the said nested prior is closest is that with hyperparameters $\boldsymbol\tau = \textbf{M}^+\boldsymbol\upsilon$ and $m=w$. Expressing the Bayes factor purely in terms of the prior and posterior normalizing constants we have
\begin{equation}\label{evi_as_norm_cons}
    \log\frac{H(\boldsymbol\tau,m)}{G(\boldsymbol\upsilon, w)}+ \log \frac{G(\textbf{M}\boldsymbol T_n+\boldsymbol\upsilon, n+w)}{H(\boldsymbol T_n +\boldsymbol\tau, n+m)}.
\end{equation}
It is not difficult to show that the optimized log-ratios in the statement of Theorem \ref{third_theorem} are positive numbers; hence the matched hyperparameters maximize the first term of (\ref{evi_as_norm_cons}), thereby mitigating the effect of the hyperparameters on the asymptotic behavior of the model selection procedure.

\section{Evidence for Random Graphs}\label{sec:graphs_er_vs_ie}

Let $G$ denote a random graph on $n_v$ vertices generated by some model identified with a symmetric edge probability matrix $\textbf{P}\in (0,1)^{n_v\times n_v}$. If we assume that $G$ is undirected and permits self-loops, then $P_{ij}$ indicates the probability that an edge exists between vertices $i$ and $j$. We wish to determine whether $G$ arose from either an Erd\"os-R\'enyi \citep{JMLR}, stochastic block- \citep{SBM-83}, random dot-product graph (RDPG) \citep{Young-07}, or an independent edge full-rank graph model. Each model posits a particular structure for $\textbf{P}$; at the extremes of this selection problem, the Erd\"os-R\'enyi (ER) model prescribes $\textbf{P} = p\boldsymbol1_{n_v}\boldsymbol1_{n_v}^\top, p \in (0,1)$, and the full-rank model assumes that $\textbf{P}$ is fully rank-$n_v$. Meanwhile, in an RDPG we assume that there exists a matrix $\textbf{X}\in \mathbbm R^{n_v\times d}$ with $d<n_v$ such that $\textbf{P} = \textbf{XX}^\top$; and in a positive definite stochastic blockmodel (SBM) we assume that such $\textbf{X}$ is structured such that its rows consist of $2 \leq K <n_v$ $d$-vectors repeated according to the block memberships of the graph's nodes.

If we are to simply identify each model with the number of parameters necessary to describe them, then we see that the ER model has only 1 parameter, rank-$d$ $K$-block (with $1\leq d \leq K$) SBMs have $dK$ parameters  (plus an additional $K-1$ identifiable-up-to-permutation block membership probability parameters when the nodes' true memberships are unknown), rank-$d$ RDPGs have $dn_v$ parameters denoting the entries of the matrix $\textbf X$, and full-rank models have $\binom{n_v}{2}+n_v$ parameters. We present the complexities of the various possible models in Figure \ref{fig:fit-complexity}. For SBMs and RDPGs, the parameters corresponding to the latent-position decomposition of $\textbf{P}$ are subject to identifiability only up to orthogonal transformation in $\mathbb{R}^d$ as described by \cite{JMLR}.

The task of model selection for such settings has largely concerned the choice of embedding dimension $d$ and number of blocks $K$ for an SBM, as the other models describe settings which are too idealistically simple (ER) or complex (RDPGs, full-rank) to be of much inferential utility. The asymptotics of the spectra of random graphs' adjacency matrices \citep{Avanti-16,Tang-18} have seemingly permitted the appropriation of tools commonly associated with Gaussian mixture models (GMMs) to this task. \cite{yang-21} performed model selection for SBMs via simultaneous estimation of $d$ and $K$, which involved fitting GMMs via the EM algorithm on embeddings of varying dimension and penalizing the observed log-likelihood with the usual BIC penalty; meanwhile, \cite{passino-2020} tackled the problem with a fully Bayesian formulation, complete with normal-inverse-Wishart joint conjugate priors on the means and covariance matrices of the components of the spectral embeddings' GMMs.

While either of these approaches improve upon previous methods - e.g., sequential selection of spectral embedding dimension and mixture complexity \textit{\`a la} \cite{zg-2006} - it is our contention that these and similar methods too heavily emphasize the asymptotic Gaussianity which arises and lose sight of the original random graph setting. The authors of \cite{Pisano-2022} argued that fitting fully parameterized GMMs to a random graph's spectral embeddings over-complicates the low-rank structure, since each component's asymptotic covariance can be explicitly written as a function of the true SBM latent positions and block membership proportions; hence, including the entries of the covariance matrices among the number of parameters to be penalized in the BIC penalty inherently over-counts the parameters of the SBM from which the data arise. We deem it similarly inappropriate, in a Bayesian context, to place priors on the component means and co-variances which are conjugate only for the complete-data formulation of the embeddings' asymptotic distributions; clearly, under such a paradigm one can place undue prior weight (however infinitesimally small) on component means and covariances which cannot possibly correspond to an SBM (e.g., $d$-variate normal priors on the component means have support outside the unit sphere in $\mathbbm R^d$). To motivate the development of an evidence-based model selection criterion for random graphs, let us consider a simplified problem to which the content of Section \ref{sec:background} may be easily applied.

\subsection{Evidence-Based Model Selection of Erd\"os-R\'enyi and Independent Edge Full-Rank Models}

Suppose $\textbf{A}$ is the adjacency matrix of a random graph $G$ (undirected, permitting self-loops) on $n_v$ vertices, and we wish to determine whether $G$ was generated by an ER$(p)$ model or an independent-edge full-rank random graph model with adjacency probability matrix $\textbf{P}$ (hereafter denoted IE$(\textbf{P})$). This problem is exactly equivalent to whether $n:=\binom{n_v}{2}+n_v$ independent Bernoulli random variables can be considered to be all identically distributed, or whether no pair is identically distributed. Moreover, this scenario is one in which BIC's unsuitability is immediately apparent, since the maximum likelihood estimator of $\textbf P$ under the IE model is exactly $\textbf A$ and thus not within the interior of the parameter space, which is one of the requirements for the appropriateness of the BIC approximation to the evidence \citep{Biernacki-1998}. In this case the BIC for the IE model for any graph on $n_v$ vertices can be shown to be exactly $-(n/2)\log n_v$.

We will first cast this problem within a general Bayesian framework before reformulating it in terms of the content of Section \ref{2.2}. As the dimensionality of the parameter-space for the IE model is exactly $n$, we will consider it as the ``full'' model. Since the $i$-th Bernoulli trial associated with the existence of the corresponding edge is a fully rank-one exponential family random variable, the adjacency parameter for that single edge may take a conjugate beta prior; the hierarchical model in this case can be expressed as
\begin{align*}
    A_{i}\ | \ p_i &\sim \text{Bern}(p_i), \ i=1,\dots, n\\
    p_i &\overset{ind.}{\sim} \text{Beta}(\alpha_i, \beta_i)
\end{align*}
with $\alpha_i,\beta_i>0$ for all $i$. Note that the edge probabilities need not be identically distributed. The posteriors for the edge probabilities can be easily determined as
\begin{equation*}
    p_i\ | \ A_i=a_i \sim \text{Beta}(\alpha_i+a_i, \beta_i+1-a_i).
\end{equation*}
The log-evidence for this full model, then, can be exactly computed as
\begin{align*}
    \log E_{IE} = &\sum_{i=1}^n \big(a_i\log p_i + (1-a_i)\log(1-p_i)\big)\\
    &- \sum_{i=1}^n\log\frac{p_i^{\alpha_i+a_i-1}(1-p_i)^{\beta_i+1-a_i-1}\Gamma(\alpha_i+\beta_i+1)}{\Gamma(\alpha_i+a_i)\Gamma(\beta_i+1-a_i)}\\
    &+\sum_{i=1}^n\log\frac{p_i^{\alpha_i-1}(1-p_i)^{\beta_i-1}\Gamma(\alpha_i+\beta_i)}{\Gamma(\alpha_i)\Gamma(\beta_i)}.
\end{align*}
Simplification yields this as exactly equal to
\begin{equation*}
    \log E_{IE} = \sum_{i=1}^n a_i\log\frac{\alpha_i}{\alpha_i+\beta_i} + \sum_{i=1}^n (1-a_i)\log\frac{\beta_i}{\alpha_i+\beta_i},
\end{equation*}
precisely the log-likelihood of a specific collection of $n$ independent Bernoulli trials with respective success probabilities $\alpha_i/(\alpha_i+\beta_i)$.

Under the ER$(p)$ formulation we have that the probability of each edge existing is governed by a single parameter $p$; the ensuing conjugate Bayesian formulation is
\begin{align*}
    A_i \ | \ p &\overset{i.i.d.}{\sim} \text{Bern}(p), i = 1,\dots, n\\
    p &\sim \text{Beta}(\alpha, \beta)
\end{align*}
with posterior
\begin{align*}
    p \ |\ A_1=a_1,\dots, A_n=a_n \sim \text{Beta}\big(\alpha + \sum_{i=1}^na_i, \beta+n-\sum_{i=1}^n a_i\big),
\end{align*}
and log-evidence
\begin{align*}
    \log E_{ER} &= \log\frac{\text{B}(\alpha+\sum_{i=1}^na_i, \beta+n-\sum_{i=1}^na_i)}{\text{B}(\alpha, \beta)}\\
    &= \log\frac{(\alpha)_{(\sum_{i=1}^na_i)} (\beta)_{(n-\sum_{i=1}^na_i)}}{(\alpha+\beta)_{(n)}},
\end{align*}
where the expressions in the second line on the right-hand-side are Pochhammer symbols. For the special case where $\alpha=\beta=1$ the log-evidence simplifies to
\begin{equation*}
    \log E_{ER} = -\log\bigg[ (n+1) \binom{n}{\sum_{i=1}^na_i}\bigg].
\end{equation*}

Although we have the explicit expression of the log-evidence of ER($p$), some components in the expression may be too computationally expensive to compute for even moderately-sized graphs. Hence, when we compute the evidence in practice, we instead leverage the relationship among evidence, likelihood, prior, and posterior, to estimate the log-evidence of ER($p$):
\begin{equation*}
\log E_{ER} = \log f_{\mathbf{A}|p^*}(\mathbf{A}) - \log \rho_{p|\mathbf{A}}(p^*) + \log \rho_{p}(p^*)
\end{equation*}
where $p^*$ is the MAP solution, and by simple computation, equals $\min(\frac{\alpha+s-1}{\alpha+\beta+n-2},1)$, where $\alpha=\beta$ are the Beta prior parameters of ER($p$), $n=\binom{n_v}{2}+n_v$ and $s=\sum^n_{i=1}a_i$.

Our work in Section \ref{sec:background} was motivated by the problem of hyperparameterizing nested exponential family models, since regardless of the graph observed there exist hyperparameters which lead to an evidence-based conclusion of either model over the other. To see this for the present problem, suppose the shape hyperparameters in the ER model are both 1 as in the simplified case --- which leads to that model's log-evidence only depending on $n$ and $\sum_{i=1}^n a_i$ --- and also suppose that all the $p_i$ are identically distributed $\text{Beta}(\alpha, \beta)$ \textit{a priori} in the IE model. The log-evidence of the latter, then, is a function of the number of edges present in the graph and the value of $\lambda := \alpha/(\alpha+\beta) \in (0,1)$. The Bayes factor for these two models is thus
\begin{equation*}
    \log \frac{E_{IE}}{E_{ER}} = \sum_{i=1}^n a_i\log\lambda + (n-\sum_{i=1}^n a_i)\log(1-\lambda) + \log\bigg[ (n+1) \binom{n}{\sum_{i=1}^n a_i}\bigg].
\end{equation*}
This expression is clearly concave on $\lambda\in (0,1)$ and achieves its maximum at $\lambda = \sum_{i=1}^n a_i/n$. For $n$ large we can invoke Stirling's approximation $n!\approx \sqrt{2\pi n}(n/e)^n$ to determine that the Bayes factor evaluated at this particular value of $\lambda$ is $O(\log n)$, and thus positive, which leads to us selecting model IE. Likewise, if we evaluate the Bayes factor at $\lambda$ close enough to either 0 or 1 we can blow up either of the first two terms to render at least one of them negative enough to dominate the third term, which leads to us selecting model ER. Clearly the hyperparameterization matters here.

\subsection{Models IE and ER as Canonical Exponential Families}\label{3.2}

To ``Occamize'' this model selection problem, we shall re-characterize the IE likelihood as a fully rank-$n$ exponential family in canonical form, and the ER model as a rank-$1$ linear submodel thereof. Following this we will derive appropriate conjugate priors for the latter model based around that of the former.

Letting $\boldsymbol a\in\{0,1\}^n$ denote the realized vector of edge existences, the likelihood for the general IE model is
\begin{align*}
    f(\boldsymbol a \ | \ \boldsymbol p) &= \prod_{i=1}^n p_i^{a_i}(1-p_i)^{1-a_i}\\
    &= \exp\left(\sum_{i=1}^n a_i\log{\frac{p_i}{1-p_i}}+\sum_{i=1}^n \log{(1-p_i)} \right),
\end{align*}
where $\boldsymbol p$ denotes the $n$ potentially unique entries of $\textbf{P}$. Defining $\boldsymbol\theta$ such that $\theta_i = \log(p_i/(1-p_i))$, and ${\mathcal{A}}(\boldsymbol\theta) = \sum_{i=1}^n\log(1+\exp\theta_i)$ we observe that the likelihood can be rewritten as
\begin{equation*}
    f(\boldsymbol a \ | \ \boldsymbol \theta) = \exp\left(\boldsymbol a^\top \boldsymbol\theta - \sum_{i=1}^n\log(1+\exp\theta_i)\right);
\end{equation*}
i.e., as a fully rank-$n$ exponential family in canonical form with base measure 1, sufficient statistic $\boldsymbol a$, canonical parameter $\boldsymbol\theta$ taking values in canonical parameter space $\mathbbm R^n$, and log-partition ${\mathcal{A}}(\boldsymbol\theta)$. That the log-partition is convex can be noted from each summand being the log-partition of a Bern$(\exp(\theta_i)/(1+\exp(\theta_i)))$ density, and that the sum of convex functions is also convex.

Via the content and methods of Section \ref{2.1} we can construct a conjugate prior for the IE model by defining $m > 0$ and $\boldsymbol\tau$ in the interior of the convex support of $m\boldsymbol A$ (i.e., $(0, m)^n$), obtaining
\begin{align*}
    \rho_{IE}(\boldsymbol\theta) &\propto \exp\left(\boldsymbol\tau^\top\boldsymbol\theta - m\bigg(\sum_{i=1}^n\log(1+\exp\theta_i)\bigg)\right)\\
    &= \prod_{i=1}^n \exp\left(\tau_i\theta_i - m\log(1+\exp\theta_i)\right).
\end{align*}
If we apply the change-of-variables $p_i = \exp(\theta_i)/(1+\exp(\theta_i))$ to each term in the product, we see that this prior corresponds to the joint density of independent beta random variables with $\alpha_i = \tau_i$ and $\beta_i = m-\tau_i$.

Under the ER formulation the entries of $\boldsymbol p$ are all equal, and thus so are the entries of $\boldsymbol\theta$ (say, to some value $\eta\in \mathbbm R$). The likelihood in this case is precisely
\begin{align*}
    f(\boldsymbol a \ |\ \boldsymbol1_n\eta) &= \exp\left(\langle\boldsymbol a, \boldsymbol1_n\eta\rangle - n\log(1+\exp\eta)\right)\\
    &= \exp\left((\boldsymbol{a}^\top\boldsymbol1_n) \eta - n\log(1+\exp\eta)\right);
\end{align*}
i.e., that of an rank-1 exponential family with base measure 1, sufficient statistic $\boldsymbol{a}^\top\boldsymbol1_n = \sum_{i=1}^n a_i\in \{0,1,\dots,n\}$, canonical parameter $\eta\in \mathbbm R$, and log-partition $B(\eta) = n\log(1+\exp\eta)$. Given a hyperparameterization $(\boldsymbol\tau, m)$ for the IE canonical parameter $\boldsymbol\theta$ the first statement of Theorem \ref{third_theorem} yields the nearest information-theoretic conjugate prior for $\eta$ as
\begin{equation*}
    \rho_{ER}(\eta) \propto \exp\left((\boldsymbol\tau^\top\boldsymbol1_n)\eta- nm\log(1+\exp\eta)\right),
\end{equation*}
which, we find after applying the change-of-variables $p = \exp(\eta)(1+\exp\eta)$, corresponds to a $\text{Beta}(\boldsymbol\tau^\top\boldsymbol1_n,nm-\boldsymbol\tau^\top\boldsymbol1_n)$ prior for the ER model's single parameter.  Likewise, had we begun with a $\text{Beta}(\upsilon, w-\upsilon)$  prior for $p$ in the ER model with $w>0,\upsilon\in(0, w)$, the second statement of Theorem \ref{third_theorem} gives the nearest information-theoretic conjugate prior for the entries of $\boldsymbol p$ as the product of $n$-independent $\text{Beta}(\upsilon/n, (\upsilon-w)/n)$ densities.

As a consequence, we observe that specifying i.i.d. uniform priors on the distinct entries of $\boldsymbol p$ does not induce the same prior for $p$; indeed, such a choice corresponds to the latter being exactly $\text{Beta}(n,n)$ distributed \textit{a priori}. Conversely, specifying a uniform prior for $p$ induces independent $\text{Beta}(1/n,1/n)$ priors for the entries of $\boldsymbol p$. For the latter case we observed in the previous subsection that the Bayes factor between IE and ER depends only on the number of edges $\sum_{i=1}^n a_i$ and the common expected value of the Beta priors, which is exactly $1/2$ after hyperparameter matching; the Bayes factor is exactly
\begin{equation*}
    \log\frac{E_{IE}}{E_{ER}} = \log\frac{n+1}{2^n} + \log{\binom{n}{\sum_{i=1}^n a_i}}.
\end{equation*}
Since $2^n = \sum_{k=0}^n \binom{n}{k}$, the Bayes factor is positive when $\binom{n}{\sum_{i=1}^n a_i}$ is strictly larger than the average binomial coefficient with first argument $n$. Marginally under the full model the graph with sequence of edge existences $A_1,\dots, A_n$ has probability $(1/2)^n$ of being observed; hence the number of edges $\sum_{i=1}^n A_i$ has a binomial marginal distribution with success probability $0.5$ on $n$ trials. The Bayes factor favors IE when
\begin{equation*}
    \frac{1}{n+1} < {\binom{n}{\sum_{i=1}^n A_i}} \bigg(\frac{1}{2}\bigg)^n.
\end{equation*}

A natural question, of course, is whether the Bayes factor is even justifiable for this problem. Were we to compute a prior-corrected BIC we again run into the issue of both the log-likelihood and log-prior for IE necessarily being evaluated at the boundary of the parameter space. The log-likelihood evaluates to 0, but the log-prior evaluates to $\infty$. The evidence for ER will always be finite when the uniform prior for $p$ is used, hence the prior-corrected BIC always results in IE being selected assuming we have matched the hyperparameters of the larger model. The confounding issue here is that the dimension of the parameter space for the enveloping model is not fixed, but rather grows at $O(n_v^2)$, hence we cannot simply invoke the earlier theory pertaining to settings in which the parameter space retains the same dimension as we observe more data. We do, however, obtain the following result, which effectively completes the analysis begun in Section 5.1 of \cite{Pisano-diss}; the proof may be found in Appendix A.3.
\begin{theorem}\label{thm:BF_cons_graphs}
    Let $E^{(n_v)}_{IE}$ and $E^{(n_v)}_{ER}$ denote the evidence for the subscripted models based on a graph $G$ with $n_v$-vertices assuming we have regularized ER with $U(0,1)$ for $p$ and IE with $n$-independent $\text{Beta}(1/n,1/n)$ priors for the entries of $\boldsymbol{p}$. We have that
    \begin{equation*}
        \lim_{n_v\to\infty} P\bigg(\log \frac{E^{(n_v)}_{IE}}{E^{(n_v)}_{ER}} > 0 \ \big| \ G\sim \text{ER}(p)\bigg) = \begin{cases}
            1, &\text{if}\ p = \frac{1}{2}\\
            0, &\text{if}\ p \neq \frac{1}{2}.
        \end{cases}
    \end{equation*}
    That is, the Bayes factor for IE vs ER is consistent for ER provided that $p\neq 0.5$.
\end{theorem}

\begin{figure}
    \centering
    \includegraphics[width=4in]{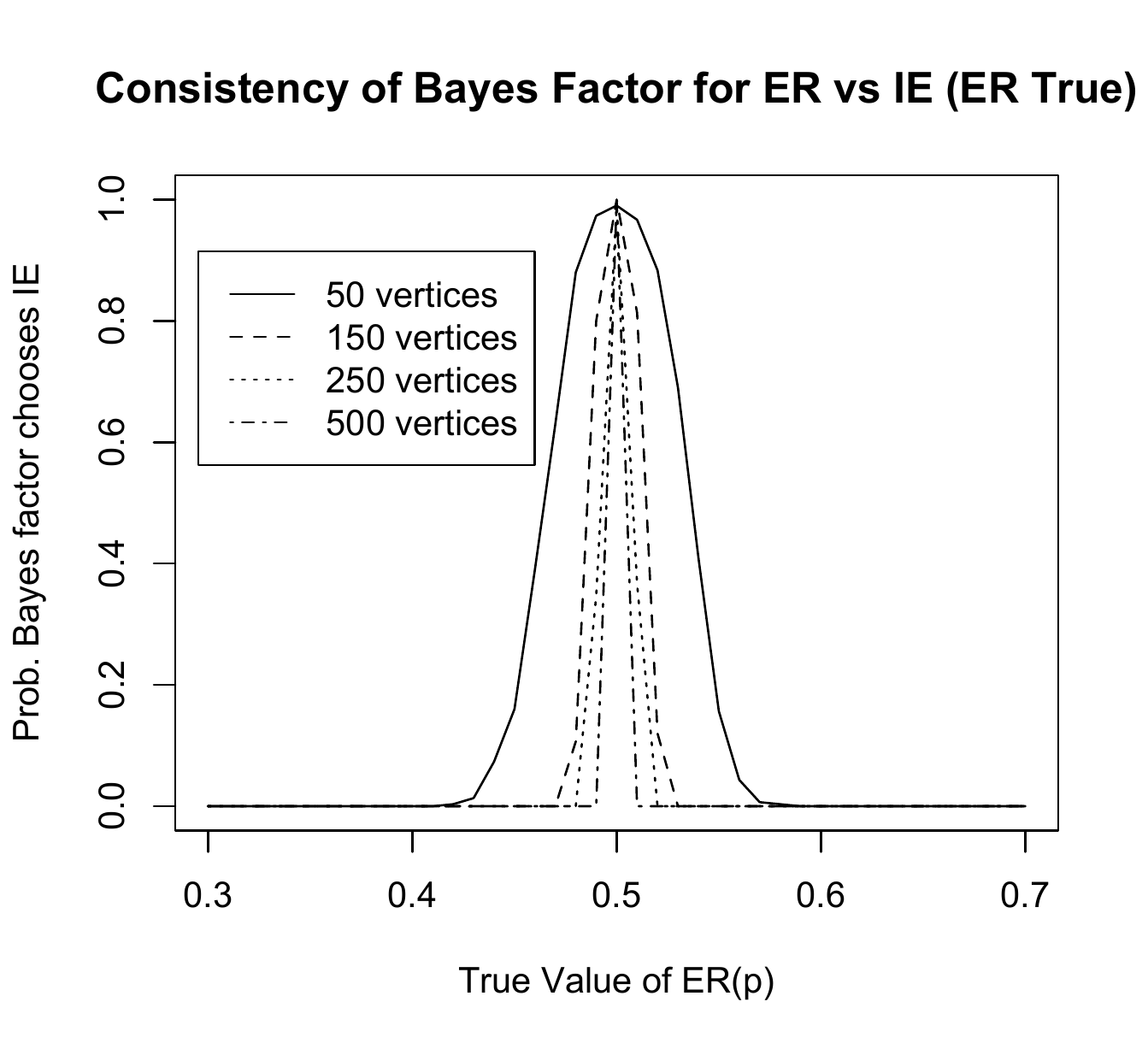}
    \caption{Prob. of the Bayes factor selecting IE when ER is true for the listed graph sizes.}
    \label{fig:graph_bayes_figure}
\end{figure}

This is to say that the Bayes factor is consistent for ER$(p)$ over IE provided that $p\neq 0.5$. The phenomenon described by this result is depicted in Figure \ref{fig:graph_bayes_figure}. For each $n_v=50, 150, 250, 500$ and $p=0.30, 0.31,\dots, 0.69,0.70$ we simulated 300 ER($p$) graphs and computed the proportion for which the Bayes factor selects the more complicated model. As $n_v$ increases, it is clear that the probability of selecting IE shrinks to 0 for all values of $p$ save near $0.5$. Increasing $n_v$ even further will yield a thinner spike about this value. This effectively corroborates what we stated above, since the Bayes factor can be guaranteed to be positive when the observed proportion of edges falls near the common prior expectation $\lambda$ of the entries of $\boldsymbol{p}$.

A natural follow-up concerns whether the Bayes factor consistently selects IE. Since the parameter space grows quadratically in the number of vertices, consistency is an ill-posed condition to aspire to; nonetheless, we can bound the probability that the Occam paradigm correctly selects IE provided that appropriate conditions on the edge probabilities hold.
\begin{theorem}\label{thm:IE_bound}
    Suppose we are in the statement of the previous theorem. Fix $\varepsilon > 3.166/\sqrt{n}$ and $\delta>0$. We have that
    \begin{equation*}
        P\bigg(\log \frac{E^{(n_v)}_{IE}}{E^{(n_v)}_{ER}} > 0 \ \big| \ G\sim \text{IE}(\boldsymbol p)\bigg) \geq P\big(Z^2 \leq ({1-n^{-\frac \delta2}})^2\log n({1-n^{-\frac \delta2}})^2\big) - \varepsilon
    \end{equation*}
    where $Z$ denotes a standard normal random variable, provided that 
    \begin{equation*}
        p_i \in \bigg(\frac{1.5830}{\varepsilon\sqrt n}, 1-\frac{1.5830}{\varepsilon\sqrt n}\bigg)
    \end{equation*}
    for all $i =1,\dots,n$ and
    \begin{equation*}
        \frac{n-\sqrt{n^{1-\delta}\log n}}{2}<\sum_{i=1}^n p<\frac{n+\sqrt{n^{1-\delta}\log n}}{2}.
    \end{equation*}
\end{theorem}

\begin{figure}[h!]
    \centering
    \includegraphics[width=4in]{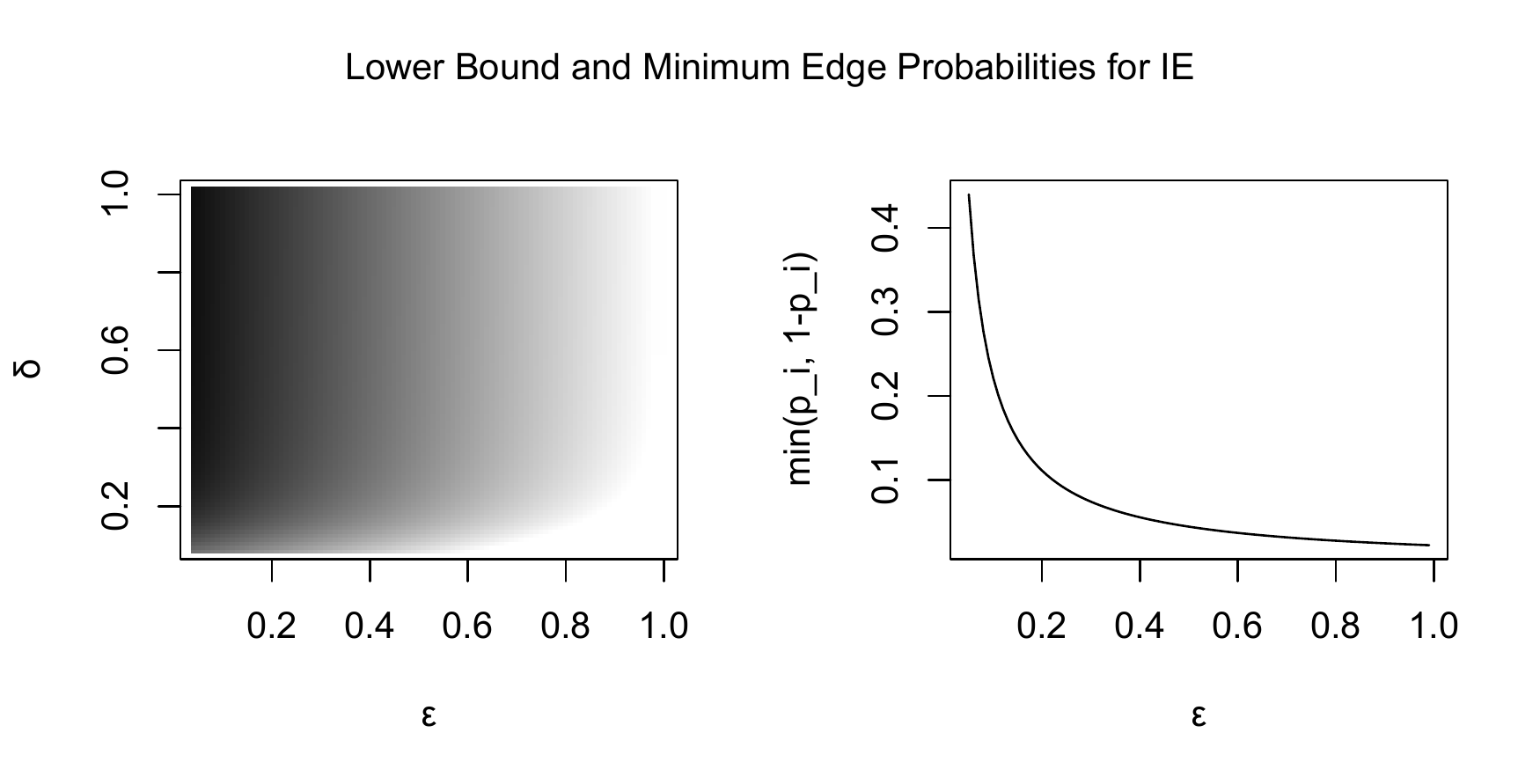}
    \caption{Left: Lower bound for the probability of evidence correctly selecting IE for $(\varepsilon, \delta)\in (0,1)^2$ when $n_v=100$. Darker regions indicate a higher lower bound. Right: Minimum edge probabilities (and complements) corresponding to various $\varepsilon.$}
    \label{fig:3.2plot}
\end{figure}

The proof of this theorem can be found in Appendix A.4. This result arises from the fact that when the graph arises from an IE$(\boldsymbol p)$ model the number of edges is the sum of $n$ independent but not identically distributed Bernoulli random variables; per \cite{Tang-2023} this sum has a Poisson binomial distribution, and thus permits a Gaussian approximation. The lower bounds for various values of $\varepsilon$ and $\delta$ and corresponding minimum edge probabilities for the case when $n_v=100$ are depicted in Figure \ref{fig:3.2plot}.

\subsection{An Approximate Occam Paradigm for a Rank-1 SBM with Known Membership}\label{sec:sbm_known}

We have, so far, only presented the task of model selection for random graphs when the only candidate models lie at the extrema of the model space. The Erd\"os-R\'enyi model is the simplest, with only one governing parameter; and the independent edge model is the most complicated, with a number of parameters of order quadratic in the number of vertices. We hereby begin the task of extending the model selection problem to encompass rank-1 SBMs.

As mentioned in the opening paragraph of the previous section, a positive definite stochastic blockmodel (SBM) model on $n_v$ vertices with $K$ (with $1\leq K< n_v$) blocks and rank $d$ ($d\leq K$) is characterized by an edge probability matrix $\textbf P = \textbf{XX}^\top$, where $\textbf{X} \in \mathbbm R^{K\times d}$. For the sake of illustration we shall primarily treat the case of a rank-1, $K$-block SBM. In practice, the vertices' block memberships must be learned or estimated (see, e.g., \citealp{Pisano-2022}); for simplicity, we will assume in this section that they are known. Such a formulation only involves $K$ parameters --- we denote it as SBM($\mathbf{X}$) with $\mathbf{X}=[X_1,X_2,\cdots,X_K]^\top$, whose block membership is known and characterized by a $K\times K$ matrix $\mathbf{B}=[X_1\ X_2\ \cdots\ X_K]^\top [X_1\ X_2\ \cdots\ X_K]$. Let vertices $1,\dots, n_1$ belong to block 1, vertices $n_{S-1}+1,\dots, n_S$ belong to block $S$ ($2\leq S\leq K$), and vertices $n_{K-1}+1,\dots, n_v$ belong to block $K$, with $1\leq n_1<n_2<\cdots<n_{K-1}<n_v$.

For illustrative purposes, let $A_{ij}$ denote the entries of adjacency matrix $\mathbf{A}$, and define
\begin{align*}
    &S_{11} := \sum_{i=1}^{n_1 }\sum_{j=i}^{n_1 } A_{ij} &O_{11}:= \sum_{i=1}^{n_1 }\sum_{j=i}^{n_1} (1-A_{ij});\\
    &S_{12} := \sum_{i=1}^{n_1}\sum_{j= n_1+1}^{n_2} A_{ij}\ &O_{12}:= \sum_{i=1}^{n_1}\sum_{j= n_1+1}^{n_2} (1-A_{ij});
\end{align*}
as well the remaining $S_{kk^{'}} \ \text{and}\ O_{kk^{'}},\ k\leq k^{'}$, with $S_{kk^{'}}=S_{k^{'}k},\ O_{kk^{'}}=O_{k^{'}k}$. Observe that $S_{kk^{'}}$ denotes the number of edges present between nodes in blocks $k$ and $k^{'}$, and $O_{kk^{'}}$ represents the number of missing edges.

The likelihood function of observing adjacency matrix $\mathbf{A}$ under this model is then
\begin{align*}
f_{\mathbf{A}|X_1,\cdots,X_K}(\mathbf{A}|x_1,\cdots,x_K) = \prod_{i=1}^K \left( x_i^{2S_{ii}+\sum_{j\neq i}S_{ij}} (1-x_i^2)^{O_{ii}} \prod_{i<j}(1-x_ix_j)^{O_{ij}}\right)
\end{align*}

Since $X_1,\cdots,X_K$ take values in $(0,1)$, we deem it natural and intuitive to \textit{a priori} assign to them independent (and not necessarily identical) beta distributions, i.e.,
\begin{align*}
X_i \sim \text{Beta}(\alpha_i,\beta_i),\ i=1,\cdots,K;
\end{align*}
Via Bayes' theorem, we obtain the posterior of $x_1,\cdots,x_K$ given $\mathbf{A}$ is
\begin{equation*}
    p_{X_1,\cdots,X_K|\mathbf{A}}(x_1,\cdots,x_K)=\frac{p_0(x_1,\cdots,x_K,\mathbf{A})}{z_p}
\end{equation*}
where
\begin{align*}
    p_0(x_1,\cdots,x_K,\mathbf{A}) = f_{\mathbf{A}|X_1,\cdots,X_K}(\mathbf{A}|x_1,\cdots,x_K)\prod^K_{i=1}\frac{x_i^{\alpha_i-1}(1-x_i)^{\beta_i-1}}{B(\alpha_i,\beta_i)}
\end{align*}
and $z_p$ is the normalization constant, which is the multi-integral of $p_0$ with respect to $dx_1,\cdots,dx_K$, and is the evidence of this SBM, denoted as $E_{SBM}$, i.e.:
\begin{equation*}
\int_0^1\cdots\int_0^1 \prod_{i=1}^K \left( x_i^{2S_{ii}+\sum_{j\neq i}S_{ij}} (1-x_i^2)^{O_{ii}} \prod_{i<j}(1-x_ix_j)^{O_{ij}}\right) \prod^K_{i=1}\frac{x_i^{\alpha_i-1}(1-x_i)^{\beta_i-1}}{B(\alpha_i,\beta_i)} dx_1\cdots dx_K
\end{equation*}

In Section \ref{3.2} we demonstrated that the assignment of a uniform prior for $p$ in ER($p$) induces independent $\text{Beta}(1/n,1/n)$ priors for the entries of $\textbf{P}$ in IE($\textbf{P}$). To obtain the induced hyperparameters for $(X_1,\cdots,X_K)$ in our SBM($\mathbf{X}$), we should optimize the Kullback-Leibler divergence for each pair, over the region of the more complex parameter space which corresponds to the simpler parameter space \citep{Klugkist-2007}. We derive that when we assign a uniform prior in ER($p$), the induced prior for SBM($\mathbf{X}$) is $X_i\sim \text{Beta}(2,1),\ i=1,\cdots,K$. We present a detailed derivation of this in Appendix A.5.

Unfortunately, the evidence of rank-1 block-$K$ SBM($\mathbf{X}$) under the induced prior $\text{Beta}(2,1)$ is a multi-integral which does not have a closed form, albeit one which may typically not be conveniently obtained; when the graph size is large, the exponents inside the integral will frequently be too large to compute the integral numerically. To obtain an approximation for the evidence, we implement the Laplace approximation, which has the form below according to \cite{MacKay2003}:
\begin{equation*}
    E_{SBM}=z_p\approx p_0(x_1^*,\cdots,x_K^*,\mathbf{A})\sqrt{\frac{(2\pi)^K}{\det J}}
\end{equation*}
where $(x_1^*,\cdots,x_K^*)$ is the maximizer of $p_0(x_1,\cdots,x_K,\mathbf{A})$, and the matrix J is defined by:
\begin{equation*}
    J_{ij}=-\frac{\partial^2}{\partial x_i \partial x_j}\ln p_0(x_1,\cdots,x_K,\mathbf{A})|_{x_1^*,\cdots,x_K^*,\mathbf{A}},\ i,j=1,\cdots,K.
\end{equation*}
To get $(x_1^*,\cdots,x_K^*)$, we use the gradient ascent algorithm; to compute the determinant of $J$, we can easily derive the first-degree derivative as
\begin{equation*}
\frac{\partial}{\partial x_i}\ln p_0 = \frac{2S_{ii}+\sum_{j\neq i}S_{ij}+\alpha_i-1}{x_i} - \frac{\beta_i-1}{1-x_i} - \frac{2O_{ii}x_i}{1-x_i^2}-\sum_{j\neq i}\frac{x_j O_{ij}}{1-x_ix_j},\ i=1,\cdots,K;
\end{equation*}
and the second derivatives
\begin{align*}
&\frac{\partial^2}{\partial x_i^2}\ln p_0 = -\frac{2S_{ii}+\sum_{j\neq i}S_{ij}+\alpha_i-1}{x_i^2} - \frac{\beta_i-1}{(1-x_i)^2} - \frac{2(1+x_i^2)O_{ii}}{(1-x_i^2)^2} - \sum_{j\neq i}\frac{x_j^2 O_{ij}}{(1-x_ix_j)^2};\\
&\frac{\partial^2}{\partial x_i \partial x_j}\ln p_0 = \frac{\partial^2}{\partial x_j \partial x_i}\ln p_0 = -\frac{O_{ij}}{(1-x_i x_j)^2},\ i,j\in\{1,2,3,4\},\ i\neq j.
\end{align*}

Although these derivatives may be straightforwardly obtained, we need to justify the use of Laplace approximation in the estimate of $E_{SBM}$. The method of Laplace approximation fits a Gaussian distribution with the mean equal to the MAP solution, so it gives an accurate estimate when the target distribution -- in our case, the posterior of SBM($\mathbf{X}$) -- is Gaussian, or close to Gaussian. Fortunately, we can invoke the Bernstein-von Mises Theorem in our case, which states that under certain regularity conditions the posterior distribution converges to a Gaussian distribution. In Appendix A.6, we show in detail that for rank-1 block-2 SBMs, the conditions of Bernstein-von Mises are satisfied, so its posterior is close to Gaussian for moderately large graphs. The conditions for the Laplace approximation for the evidence of the more general rank-1 block-$K$ SBMs can be proved similarly.

\subsection{An Approximate Occam Paradigm for a Rank-1 SBM with Unknown Membership}\label{sec:sbm_unknown}

Next we discuss a more flexible case when the block memberships are unknown. From the discussion in Section \ref{sec:sbm_known} we can see that the evidence of rank-1 block-$K$ SBM($\mathbf{X}$) depends on the number of edges within and between each block, and thus assigning nodes to different blocks would lead to different values of evidence. Therefore, when we want to estimate the evidence of rank-1 block-$K$ SBM($\mathbf{X}$) for a given graph with no knowledge of block membership, we use clustering techniques to estimate block membership given the observed adjacency matrix $\mathbf{A}$, and then approximate its evidence using the procedure in Section \ref{sec:sbm_known}.

The clustering problem for SBMs is a well-studied problems, with many extant methods in the literature \citep{mej}. In this paper we specifically advocate clustering nodes using their latent position $\mathbf{X}$, which in the rank-1 case is a length $n_v$ vector. Although $\mathbf{X}$ is not observed, we can perform clustering on the adjacency spectral embedding (ASE) $\hat{\mathbf{X}}=\mathbf{U}^{(d)}_{\mathbf{A}}(\mathbf{S}^{(d)}_{\mathbf{A}})^{\frac{1}{2}}$ where
$$(\mathbf{A}^\top\mathbf{A})^{\frac{1}{2}}=\mathbf{U_A}\mathbf{S_A}\mathbf{U}^\top_{\mathbf{A}}$$
is the spectral decomposition; $\mathbf{S}^{(d)}_{\mathbf{A}}$ is the $d\times d$ principal minor of the diagonal matrix $\mathbf{S_A}$ (which, in the case of $d=1$ is simply the largest eigenvalue of $\mathbf{A}^\top\mathbf{A}$); and the columns of $\mathbf{U}^{(d)}_{\mathbf{A}}$ are the first $d$ columns of the unitary matrix $\mathbf{U_A}$ \citep{Pisano-2022}. As \cite{JMLR} points out, $\mathbf{A}$ can be viewed as a small perturbation of $\mathbf{X}\mathbf{X}^\top$ and thus the ASE is a good estimate of $\mathbf{X}$. In fact, clustering on ASE has been proved to be effective for SBM($\mathbf{X}$): \cite{lyzinski-2014} shows that under some natural assumptions, if the non-zero eigenvalues of $\mathbf{X}\mathbf{X}^\top$ are distinct (which holds for most real-world cases), the mean square error clustering on ASE is error-free. In this way, we can determine the most likely block-$K$ assignment for any graph and then evaluate its evidence of such model.

\subsection{Model Selection for the Complete Graph}

Before we apply our evidence-based criterion to the model selection problem that includes ER($p$), IE($\mathbf{P}$), and rank-1 SBM($\mathbf{X}$), we compute and compare the evidences in the extreme case when the observed graph under consideration is complete. We obtain the following theorem; the proof of which we have relegated to Appendix A.7. Another extreme case occurs when the observed graph is empty, but we contend that it would make little sense to discuss the generative model in the presence of such data.

\begin{theorem}\label{thm:ExtremeCase}
 Let $G$ be the complete graph on $n_v$ vertices with $n$ possible edges, arisen from either ER($p$), IE($\textbf{P}$), or rank-1 SBM($\mathbf{X}$) with any block number $1<K\leq n_v$, with respective priors $\rho_{ER}=\text{Beta}(1,1)$, $\rho_{IE}=\text{Beta}(1/n,1/n)$, and $\rho_{SBM}=\text{Beta}(2,1)$. In this case of complete graph, ER is always selected by the evidence.
\end{theorem}
This result is intuitively not surprised - a complete graph is indeed very likely to come from a model with each edge probability close to 1, and thus an ER($p$) model with $p\approx 1$ seems to be a fair choice.

\section{Simulation Results}\label{sec:simulation}
In this section, we detail and present the results of simulations which demonstrate the numerical consistency of our Occam paradigm for the task of selecting from ER($p$), IE($\mathbf{P}$), and rank-1 SBM($\mathbf{X}$). For each simulated instance, we computed the evidences of these models and selected the model with the largest evidence.

\begin{figure}
    \centering
    \includegraphics[width=\linewidth]{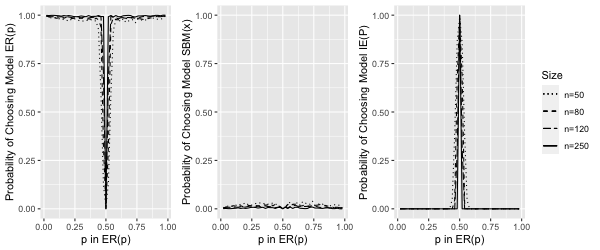}
    \caption{Probability of the criterion selecting each model when ER is true for different $n_v$.}
    \label{fig:FromER_1}
\end{figure}

We first sampled 1000 graphs from the ER($p$) model for each value of $p$ in the unit interval. Figure \ref{fig:FromER_1} shows the probability of the evidence selecting among ER($p$), rank-1 block-2 SBM($\mathbf{X}$), and IE($\mathbf{P}$), for different graph sizes. At most values of $p$, this criterion correctly chooses the true model ER($p$); when $p$ is around 0.5, the criterion chooses IE($\mathbf{P}$). As the size of graph increases, the regions in which evidence chooses IE($\mathbf{P}$) or SBM($\mathbf{X}$) shrink, indicating that the probability of successfully choosing ER($p$) goes to 1, except for the singularity at $p=0.5$, at which the criterion chooses IE($\mathbf{P}$) instead, \textit{\`a la} Theorem \ref{thm:BF_cons_graphs}.

\begin{figure}[ht!]
    \centering
    \includegraphics[width=0.5\linewidth]{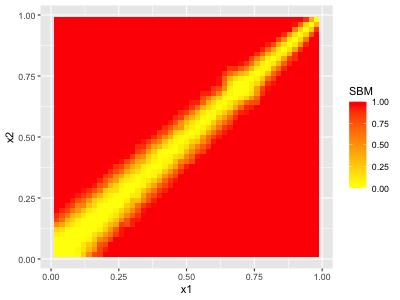}
    \caption{Heatmap of the criterion selecting SBM when SBM is true for $n_v=50$.}
    \label{fig:FromSBM_50}
\end{figure}

\begin{figure}[ht!]
    \centering
    \includegraphics[width=0.5\linewidth]{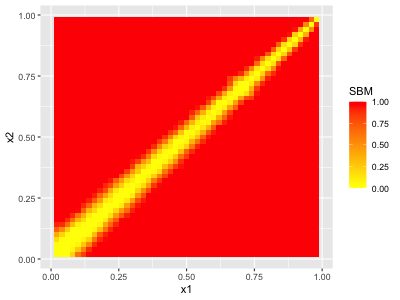}
    \caption{Heatmap of the criterion selecting SBM when SBM is true for $n_v=80$.}
    \label{fig:FromSBM_80}
\end{figure}

\begin{figure}[ht!]
    \centering
    \includegraphics[width=0.5\linewidth]{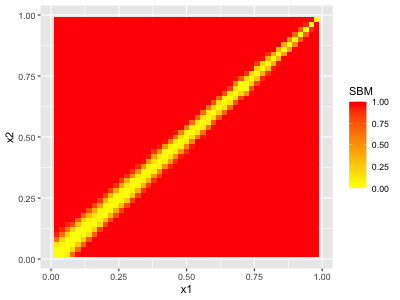}
    \caption{Heatmap of the criterion selecting SBM when SBM is true for $n_v=120$.}
    \label{fig:FromSBM_120}
\end{figure}

\begin{figure}[ht!]
    \centering
    \includegraphics[width=0.5\linewidth]{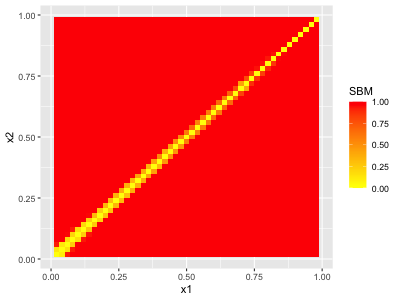}
    \caption{Heatmap of the criterion selecting SBM when SBM is true for $n_v=250$.}
    \label{fig:FromSBM_250}
\end{figure}

We also sampled 100 graphs from a rank-1 block-2 SBM($\mathbf{X}$) model for each pair of $(x_1,x_2)\in (0,1)^2$. Figure \ref{fig:FromSBM_50} to Figure \ref{fig:FromSBM_250} show the empirical proportion of times (in red) evidence selected this SBM($\mathbf{X}$), the correct model. We can see that at most values of $(x_1,x_2)$, this criterion successfully chooses the true model, except when $x_1$ and $x_2$ are close. This intuitively makes sense because when $x_1=x_2$, the rank-1 block-2 SBM model is basically an ER model. Also, the regions of mistakenly choosing other models shrink as the size of graph increases, indicating that the probability of successfully choosing the true model SBM($\mathbf{X}$) appears to go to 1 except for the singularity line $x_1=x_2$.

\begin{figure}[ht!]
    \centering
    \includegraphics[width=0.7\linewidth]{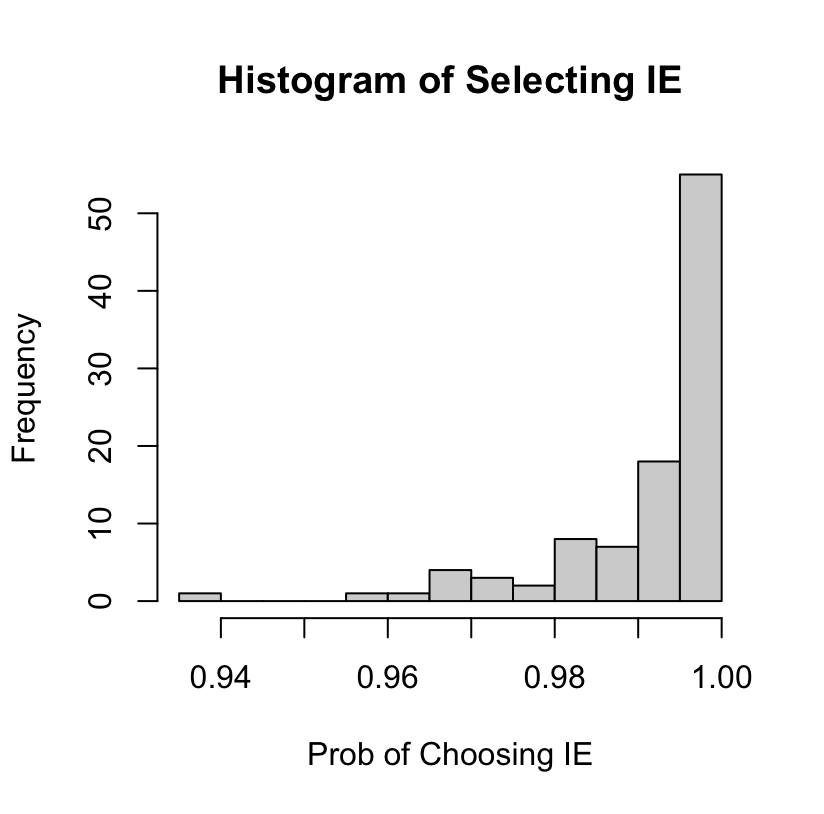}
    \caption{Histogram of the rates of correctly selecting IE when IE is true.}
    \label{fig:ie_histogram}
\end{figure}

We then explored how the evidence-based criterion works when the true model is IE($\mathbf{P}$). We randomly sampled a symmetric $50\times 50$ probability matrix $\mathbf{P}$, whose entries are independently drawn from a uniform distribution. Then we sampled 1000 graphs from this IE($\mathbf{P}$) model, and computed the evidences of ER($p$), rank-1 block-2 SBM($\mathbf{X}$), and IE($\mathbf{P}$). We then calculated the proportion of times our criterion correctly selected IE($\mathbf{P}$). We repeated this procedure 100 times, and we found that the average success rate is 99.16\%, and the lowest success rate is 93.7\%. Figure \ref{fig:ie_histogram} is the histogram of the success rates, which shows the empirical consistency of this criterion when the true model is IE($\mathbf{P}$).

\section{Model Selection on Real Data}\label{sec:application}

To demonstrate the practicality of the Occam paradigm, we apply it to connectome data analyzed by \cite{yang-21}, consisting of 112 adjacency matrices representing the connectivity of scanned human brains. The vertices, representing either individual neurons or regions of alike neurons, can be categorized into two blocks, one representing white matter, and the other representing gray matter. Alternatively, the vertices can be categorized into two blocks representing the left and right hemispheres. Hence, we speculate that a two-block SBM (either blocks of white/gray or blocks of left/right), or a four-block SBM (white/left, white/right, gray/left, gray/right) may be suitable for such data.

Since self-loops are not allowed among these data, we slightly modify our formulation to account for them. The number of possible edges becomes $n^*=\binom{n_v}{2}$ instead of $n=\binom{n_v}{2}+n_v$, and the number of observed edges becomes $s^*=\sum\limits_{i=1}^{n^*}a_i$.

For each graph, we next compute its evidence under five different models: ER($p$), IE($\mathbf{P}$), rank-1 block-2(white/gray) SBM($\mathbf{X}$) with $\mathbf{X}=[X_1,X_2]$, rank-1 block-2(left/right) SBM($\mathbf{X}$) with $\mathbf{X}=[X_1,X_2]$, and rank-1 block-4 SBM($\mathbf{X}$) with $\mathbf{X}=[X_1,X_2,X_3,X_4]$. Among 112 graphs, 45 of them are identified as two-block SBM of white/gray by our criterion, and 67 of them are identified as four-block SBM. Table \ref{table1} shows the five-number summaries for the log-evidences for all 112 graphs. Indeed, the evidences of 2-block SBM of white/gray and 4-block SBM are very close, and are larger than the evidences of other models. It indicates that there is a clear pattern of blocks for white matter and gray matter, while the neurons in the left hemispheres do not have significant connectivity difference from the neurons in the right hemispheres.

\begin{table}[h!]
\begin{center}
\scalebox{0.8}{
\begin{tabular}{l|lllll}
    & Minimum & First Quartile & Median & Third Quartile & Maximum \\ \hline
$\log E_{ER}$ & -65089.72 & -56933.92 & -54620.67 & -52564.26 & -47053.30 \\
$\log E_{SBM2,\ white/gray}$ & -59730.56 & -53349.17 & -51279.04 & -49471.83 & -44492.69 \\
$\log E_{SBM2,\ left/right}$ & -65025.19 & -56887.11 & -54621.95 & -52515.00 & -47013.32 \\
$\log E_{SBM4}$ & -59701.28 & -53349.78 & -51272.00 & -49458.50 & -44432.01 \\
$\log E_{IE}$ & -203619.6 & -202558.4 & -202028.8 & -200971.8 & -198341.3 \\     
\end{tabular}}
\caption{Five-number summaries for the log-evidences for all 112 brain scans.}
\label{table1}
\end{center}
\end{table}

\section{Discussion}\label{sec:discussion}
We have proposed and presented the application of the Occam paradigm to random graph models. In summary, our paradigm involves computing
the marginal likelihood (i.e., Bayesian evidence) of each candidate model for an observed graph, and results in the one with the largest evidence being selected. We have demonstrated the theoretical and empirical consistency of the evidence for the task of determining an appropriate generative model for network data, between the simplest ER model and the most complex IE model. We next extend the framework to include the rank-1 block-$K$ stochastic blockmodel, which presents itself as a more useful model than ER and IE for network data. By evaluating the evidences of these models for simulated graphs, we have demonstrated the consistency of this evidence-based criterion for such data. We then readily applied the analysis to some connectome data, whose vertices can be categorized into two blocks or four blocks; the SBMs were indeed selected as the most appropriate models.

A natural limitation of the Occam paradigm is that it only includes a few models. We plan to fill out the ``complexity gap'' depicted in Figure \ref{fig:fit-complexity} and extend the framework to include more general graph models. In Section \ref{sec:sbm_known} when we use Laplace approximation to evaluate the evidence of rank-1 SBM, the approximation is only valid when the number of edges in each block is sufficiently large so that Bernstein-von Mises can be invoked. Therefore, we cannot apply the same method to rank-1 RDPG($\mathbf{X}$), which is a rank-1 SBM with $n_v$ blocks, each containing only one node. One possible approach is to evaluate the evidence of rank-1 RDPG($\mathbf{X}$) using the evidence of rank-1 SBM($\mathbf{X}$) with the maximal block number $K$ such that Laplace approximation can be used --- as \cite{wolfe-2013} point out, RDPG($\mathbf{X}$) can be well-approximated by a SBM($\mathbf{X}$) with a sufficiently large number of blocks. However, this approach is only for approximation, and we aim to evaluate the exact evidence of rank-1 RDPG($\mathbf{X}$) in the future.

In addition, for now we have only considered rank-1 SBMs, and to include SBMs of rank $d>1$, placing beta priors on each latent position, while seemingly natural, will run into the issue of identifiability of the latent positions only up to orthogonal transformation \citep{JMLR}. Inspired by \cite{Pisano-diss}, one can express the latent positions of a model's block probability matrix $\mathbf{B}$ in polar coordinates and then assign beta priors on this characterization of the latent positions. For example, in the 2-rank $K$-block SBM with known membership, the entries of $\mathbf{B}$ can be written as $B_{kj}=\sqrt{R_k R_j}\cos{(\theta_k-\theta_j)}$, where $R_k\sim U(0,1)$ and $\theta_k\sim U(0,\pi/2)$. We hope to extend this prior to the case of higher dimensions and then approximate the posterior densities via MCMC. After accomplishing this, we will be able to include higher-dimensional SBMs in our exploration of evidence applied to random graphs.

Furthermore, one may argue that using clustering to determine the membership of SBM($\mathbf{X}$) in Section \ref{sec:sbm_unknown} might be inaccurate in some cases, so eventually we would like to extend the Occam paradigm to settings which involve missing data. As in \cite{Pisano-diss}, we will begin with the $K$-component $d$-dimensional Bayesian Gaussian mixture model. We can then adapt the integrated classification likelihood (ICL) proposed by \cite{Biernacki-1998} and determine its behavior and performance when the component variance-covariance matrices shrink in the number of observations. This scenario is relevant to the Occamization of SBMs with unknown membership since the curved-Gaussian limiting results  governing low-rank representations of the adjacency matrix possess shrinking covariance matrices; for larger graphs the ensuing missing-data entropy model selection criteria would be negligible \textit{\`a la} \cite{lyzinski-2014}. Future work will involve the direct application of the missing-data Occam paradigm to random graphs.

A natural objection to the Occam paradigm which a classical statistician may offer is the necessity of a proper prior for the simplest model to ensure probabilistic interpretation of the various models' evidences. We contend, however, that this concern, while valid \textit{\`a la} \cite{edwards-84}'s should not affect the model selection procedure for random graphs in practice. The uniform prior for ER imposes a prior sample size of $2$ possible edges, with $1$ observed, and the encompassing priors method imposes this on the prior for every other model being considered. Since network data tend to be large in practice (e.g., the smallest connectome in Section \ref{sec:application} had 757 vertices, with $\binom{757}{2}= 286146$ possible edges) the volume of actual observations (i.e., absent and present edges) should greatly outweigh the volume of pseudo-observations in the priors.

\section*{Acknowledgements}
The authors would like to thank Professor Avanti Athreya for her excellent comments and questions which formed the backbone of much of Section \ref{sec:discussion}.

\bibliographystyle{agsm.bst}

\bibliography{ockham_citations}
\appendix

\section{Proofs of Theorems}
\subsection{Proof of Theorem 2.1}
Define $\boldsymbol T_n = \sum_{i=1}^n\boldsymbol T(x_i)$. In the setting we describe we have
\begin{equation*}
    \mathcal F(\boldsymbol{\hat\theta}_n, \boldsymbol{x}) = \log \frac{H\big(\boldsymbol T_n + \boldsymbol\tau, n+m\big)}{H(\boldsymbol\tau, m)} + \langle \boldsymbol{\hat\theta}_n, \boldsymbol T_n\rangle - nA(\boldsymbol{\hat\theta_n}).
\end{equation*}
Since $\rho$ is a density function we have that
\begin{equation*}
    H(\boldsymbol T_n + \boldsymbol\tau, n+m) = \bigg(\int_{\boldsymbol\Theta}\exp{\{\langle \boldsymbol{\theta}, \boldsymbol T_n + \boldsymbol\tau\rangle - (n+m)A(\boldsymbol{\theta}\}}d\boldsymbol\theta\bigg)^{-1}.
\end{equation*}
We thus rewrite the flexibility as
\begin{align*}
    \mathcal F(\boldsymbol{\hat\theta}_n, \boldsymbol{x}) = \ &-\log \int_{\boldsymbol\Theta}\exp{\{\langle \boldsymbol{\theta}, \boldsymbol T_n + \boldsymbol\tau\rangle - (n+m)A(\boldsymbol{\theta}\}}d\boldsymbol\theta
     -\log\exp{\{-\langle \boldsymbol{\hat\theta}_n, \boldsymbol T_n\rangle+nA(\boldsymbol{\hat\theta_n})\}}\\
     &- \log H(\boldsymbol\tau, m).
\end{align*}

We absorb the log-exp term into the integral and obtain (upon temporarily ignoring the prior integration constant as well as the log) the integral
\begin{equation*}
    \int_{\boldsymbol\Theta}\exp{\{\langle \boldsymbol{\theta}-\boldsymbol{\hat\theta}_n, \boldsymbol T_n\rangle +\langle\boldsymbol\theta, \boldsymbol\tau\rangle - n(A(\boldsymbol\theta) -A(\boldsymbol{\hat\theta}_n))-mA(\boldsymbol{\theta}\}}d\boldsymbol\theta.
\end{equation*}
The exponent is maximized at the MAP estimate $\boldsymbol\theta = \boldsymbol{\hat\theta}_n$, so Laplace's method yields the approximation
\begin{align*}
    \int_{\boldsymbol\Theta}\exp \bigg\{ &\langle\boldsymbol{\hat\theta}_n, \boldsymbol\tau\rangle -mA(\boldsymbol{\hat\theta}_n) +\big\langle \boldsymbol T_n + \boldsymbol\tau - (n+m)\overset{\boldsymbol\cdot}A(\boldsymbol{\hat\theta}_n), \boldsymbol\theta-\boldsymbol{\hat\theta}_n\big\rangle \\
    &-\frac{1}{2}\big\langle \boldsymbol\theta-\boldsymbol{\hat\theta}_n, (n+m)\overset{\boldsymbol{\cdot\cdot}}A(\boldsymbol{\hat\theta}_n)(\boldsymbol\theta-\boldsymbol{\hat\theta}_n\big)\big\rangle\bigg\} d\boldsymbol\theta.
\end{align*}
The conditions on $\boldsymbol\tau$ and $m$ ensure that  $\boldsymbol{\hat\theta}_n = \dot A^{-1}(\frac{\boldsymbol T_n +\boldsymbol\tau}{n+m})$ exists uniquely in the interior of $\boldsymbol\Theta$. We see immediately that the second inner product vanishes as its first argument equals $\boldsymbol 0$, hence we are left with
\begin{align*}
    &\exp{\{\langle\boldsymbol{\hat\theta}_n, \boldsymbol\tau\rangle -mA(\boldsymbol{\hat\theta}_n)\}}\int_{\boldsymbol\Theta}\exp\bigg\{-\frac{1}{2}\big\langle \boldsymbol\theta-\boldsymbol{\hat\theta}_n, (n+m)\overset{\boldsymbol{\cdot\cdot}}A(\boldsymbol{\hat\theta}_n)(\boldsymbol\theta-\boldsymbol{\hat\theta}_n\big)\big\rangle\bigg\} d\boldsymbol\theta\\
    &=\exp{\{\langle\boldsymbol{\hat\theta}_n, \boldsymbol\tau\rangle -mA(\boldsymbol{\hat\theta}_n)\}}(2\pi)^{\frac{k}{2}}|\overset{\boldsymbol{\cdot\cdot}}A(\boldsymbol{\hat\theta}_n)|^{-\frac{1}{2}}(n+m)^{-\frac{k}{2}}.
\end{align*}

Flexibility is thus approximately
\begin{align*}
    -\log &\bigg(\exp{\{\langle\boldsymbol{\hat\theta}_n, \boldsymbol\tau\rangle -mA(\boldsymbol{\hat\theta}_n)\}}(2\pi)^{\frac{k}{2}}|\overset{\boldsymbol{\cdot\cdot}}A(\boldsymbol{\hat\theta}_n)|^{-\frac{1}{2}}(n+m)^{-\frac{k}{2}}\bigg) -\log H(\boldsymbol\tau, m)\\
    &= -\log\rho(\boldsymbol{\hat\theta_n}) + \frac{1}{2}\log\bigg|\frac{\overset{\boldsymbol{\cdot\cdot}}A(\boldsymbol{\hat\theta}_n)}{2\pi}\bigg| + \frac{k}{2}\log(n+m)
\end{align*}
and the difference between flexibility and the BIC penalty may be approximated as
\begin{equation*}
    -\log\rho(\boldsymbol{\hat\theta_n}) + \frac{1}{2}\log\bigg|\frac{\overset{\boldsymbol{\cdot\cdot}}A(\boldsymbol{\hat\theta}_n)}{2\pi}\bigg| + \frac{k}{2}\log(n+m) - \frac{k}{2}\log n.
\end{equation*}
Since $\boldsymbol{\hat\theta}_n\overset{p}\to \boldsymbol\theta_0$ and $\frac{n+m}{n}\to 1$, by the continuous mapping theorem we have that this expression converges in probability to
\begin{equation*}
    -\log\rho(\boldsymbol\theta_0) + \frac{1}{2}\log\bigg|\frac{\overset{\boldsymbol{\cdot\cdot}}A(\boldsymbol\theta_0)}{2\pi}\bigg|
\end{equation*}
as desired.
\qed

\subsection{Proof of Theorem 2.2}

We begin with the first minimization problem. Jensen's inequality yields
\begin{align*}
    -\mathbbm E_N \bigg\lbrack\log\frac{\rho_F(\textbf{M}^\top\boldsymbol\eta)}{\rho_N(\boldsymbol\eta)}\bigg\rbrack &\geq -\log \int_{\boldsymbol{\mathcal E}}\bigg(\frac{\rho_F(\textbf{M}^\top\boldsymbol\eta)}{\rho_N(\boldsymbol\eta)}\bigg)\rho_N(\boldsymbol\eta)d\boldsymbol\eta\\
    &= -\log\int_{\boldsymbol{\mathcal E}}\rho_F(\textbf{M}^\top\boldsymbol\eta)d\boldsymbol\eta\\
    &= -\log \bigg\lbrack H(\boldsymbol\tau, m)\int_{\boldsymbol{\mathcal E}}\exp\{\langle\boldsymbol\tau,\textbf{M}^\top\boldsymbol\eta\rangle - mA(\textbf{M}^\top\boldsymbol\eta)\}d\boldsymbol\eta\bigg\rbrack\\
    &= -\log\bigg\lbrack\frac{H(\boldsymbol\tau,m)}{G(\textbf{M}\boldsymbol\tau, m)}\int_{\boldsymbol{\mathcal E}}G(\textbf{M}\boldsymbol\tau, m)\exp\{\langle\boldsymbol\tau,\textbf{M}^\top\boldsymbol\eta\rangle - mA(\textbf{M}^\top\boldsymbol\eta)\}d\boldsymbol\eta\bigg\rbrack.
\end{align*}
The integrand in the final line is the density of the nested prior parameterized by $\boldsymbol\upsilon = \textbf{M}\boldsymbol\tau$ and $w=m$, and thus integrates over $\boldsymbol{\mathcal E}$ to 1; hence we have
\begin{equation*}
    -\mathbbm E_N \bigg\lbrack\log\frac{\rho_F(\textbf{M}^\top\boldsymbol\eta)}{\rho_N(\boldsymbol\eta)}\bigg\rbrack \geq -\log\frac{H(\boldsymbol\tau,m)}{G(\textbf{M}\boldsymbol\tau, m)} = \log\frac{G(\textbf{M}\boldsymbol\tau, m)}{H(\boldsymbol\tau, m)}.
\end{equation*}
Writing out the expectation, we have
\begin{equation*}
    -\log\frac{H(\boldsymbol\tau,m)}{G(\boldsymbol\upsilon, w)} - \langle\boldsymbol\tau,\textbf{M}^\top\mathbbm E_N\lbrack\boldsymbol\eta\rbrack\rangle + \langle\boldsymbol\upsilon,\mathbbm E_N\lbrack\boldsymbol\eta\rbrack\rangle +(m-w)\mathbbm E_N\lbrack A(\textbf{M}^\top\boldsymbol\eta)\rbrack.
\end{equation*}
When $\boldsymbol\upsilon = \textbf{M}\boldsymbol\tau$ and $w=m$ all but the first term cancel out. Hence this hyperparameterization achieves the lower bound.

We now turn to the second minimization problem, which is equivalent to solving
\begin{align*}
    &\underset{\boldsymbol\tau}{\max} \log H(\boldsymbol\tau, w)\\
    &\text{subject to}\ \textbf{M}\boldsymbol\tau = \boldsymbol\upsilon.
\end{align*}
From the definition of the maximand one can quite easily demonstrate its concavity in $\boldsymbol\tau$ via H\"older's inequality; the constrained maximization problem may thus be solved via elementary differentiation and Lagrangian multipliers.

Since $\textbf{M} \in \mathbbm R^{\ell\times k}$ is rank-$\ell$, its (right) Moore-Penrose psuedoinverse (e.g., \citealp{Horn-Johnson}, Problem 7.3.P7) may be explicitly written as
\begin{equation*}
    \textbf{M}^+ = \textbf{M}^\top(\textbf{M}\textbf{M}^\top)^{-1}.
\end{equation*}
One easily notes that $\textbf{M}\textbf{M}^+ = \boldsymbol I_\ell$. We can simplify the maximization problem further, by noting that for every $\boldsymbol\tau\in \mathcal T(\boldsymbol\upsilon)$ there exists $\boldsymbol\omega\in \text{null}(\textbf{M})$ such that $\boldsymbol\tau = \textbf{M}^+\boldsymbol\upsilon + \boldsymbol\omega$. Hence, the problem may once again be rewritten as
\begin{align*}
    &\underset{\boldsymbol\omega}{\max} \log H(\textbf{M}^+\boldsymbol\upsilon + \boldsymbol\omega, w)\\
    &\text{subject to} \ \textbf{M}\boldsymbol\omega = \boldsymbol 0.
\end{align*}

Let $\boldsymbol\lambda \in \mathbbm R^{\ell}$ denote the vector of Lagrangian multipliers. The objective function for the constrained optimization problem is
\begin{equation}\label{2.3 objective}
    \mathcal O(\boldsymbol\omega,\boldsymbol\lambda) = \log H(\textbf{M}^+\boldsymbol\upsilon + \boldsymbol\omega, w) - \boldsymbol\lambda^\top\textbf{M}\boldsymbol\omega.
\end{equation}
Elementary differentiation of (\ref{2.3 objective}) with respect to $\boldsymbol\omega$ and $\boldsymbol\lambda$ yields
\begin{align*}
    \frac{\partial \mathcal O(\boldsymbol\omega,\boldsymbol\lambda)}{\partial\boldsymbol\omega} &= \frac{1}{H(\textbf{M}^+\boldsymbol\upsilon + \boldsymbol\omega, w)}\frac{\partial  H(\textbf{M}^+\boldsymbol\upsilon + \boldsymbol\omega, w)}{\partial\boldsymbol\omega} - \textbf{M}^\top\boldsymbol\lambda\\
    \frac{\partial \mathcal O(\boldsymbol\omega,\boldsymbol\lambda)}{\partial\boldsymbol\lambda} &= -\textbf{M}\boldsymbol\omega.
\end{align*}
Setting both gradients equal to $\boldsymbol 0$, one observes that a possible solution to the second equation is $\boldsymbol\omega = \boldsymbol 0$. By once again invoking the definition of $H$ one can show
\begin{align*}
    \frac{\partial  H(\textbf{M}^+\boldsymbol\upsilon + \boldsymbol\omega, w)}{\partial\boldsymbol\omega} &= - H(\textbf{M}^+\boldsymbol\upsilon + \boldsymbol\omega, w)^2\frac{\partial}{\partial\boldsymbol\omega}\int_{\boldsymbol\Theta}\exp\{\langle\textbf{M}^+\boldsymbol\upsilon,\boldsymbol\theta\rangle - wA(\boldsymbol\theta)\}\cdot\exp\{\langle\boldsymbol\omega,\boldsymbol\theta\rangle\}d\boldsymbol\theta\\
    &= - \frac{H(\textbf{M}^+\boldsymbol\upsilon + \boldsymbol\omega, w)^2}{H(\textbf{M}^+\boldsymbol\upsilon, w)}\frac{\partial}{\partial\boldsymbol\omega}\mathbbm E_F\lbrack \exp\{\langle\boldsymbol\omega,\boldsymbol\theta\rangle\}\rbrack,
\end{align*}
where the expectation is with respect to $\rho_F$ parameterized by $\textbf{M}^+\boldsymbol\upsilon$ and $w$. Note that this expectation is precisely this distribution's moment generating function. With this in mind, we have that
\begin{equation*}
    \frac{\partial \mathcal O(\boldsymbol\omega,\boldsymbol\lambda)}{\partial\boldsymbol\omega}\bigg|_{\boldsymbol\omega = \boldsymbol 0} = -\mathbbm E_F\lbrack\boldsymbol\theta\rbrack - \textbf{M}^\top\boldsymbol\lambda
\end{equation*}
Setting this equal to $\boldsymbol 0$ and solving for $\boldsymbol\lambda$, we obtain
\begin{equation*}
    \boldsymbol\lambda = -(\textbf{M}^+)^\top \mathbbm E_F\lbrack\boldsymbol\theta\rbrack.
\end{equation*}

Due to the concavity of the original optimization problem, for all $\boldsymbol\tau \in \mathcal T(\boldsymbol\upsilon)$ we have
\begin{equation*}
    \log H(\boldsymbol\tau, w) \leq \log H(\textbf{M}^+\boldsymbol\upsilon, w).
\end{equation*}
and consequently
\begin{equation*}
    -\mathbbm E_N\bigg\lbrack\log\frac{\rho_F(\textbf{M}^\top\boldsymbol\eta)}{\rho_N(\boldsymbol\eta)}\bigg\rbrack = \log\frac{G(\boldsymbol\upsilon,w)}{H(\boldsymbol\tau, w)} \geq \log\frac{G(\boldsymbol\upsilon,w)}{H(\textbf{M}^+\boldsymbol\upsilon, w)}.
\end{equation*}
Taking $\boldsymbol\tau = \textbf{M}^+\boldsymbol\upsilon$, this lower bound is achieved.
\qed

\subsection{Proof of Theorem 3.1}

We provide an appeal using an asymptotic approximation of binomial coefficients as well as the normal approximation to the binomial distribution. Let $S:=\sum_{i=1}^n A_i$; it is clear that if $G\sim ER(p)$ then $S\sim\text{Bin}(n, p)$, where $n = \binom{n_v}{2}+n_v$. From above we know that the limitand may be rewritten as
\begin{equation*}
    P\bigg(\frac{1}{n+1} < \binom{n}{S}\bigg(\frac{1}{2}\bigg)^n\ \big|\ p\bigg).
\end{equation*}
For sufficiently large $n$ the event in which we are interested occurs when the probability mass function of a Bin$(n, \frac{1}{2})$ random variable exceeds $\frac{1}{n+1}$; since such probability mass function concentrates about $\frac{n}{2}$, we can intuit heuristically that the desired event occurs when $S\approx \frac{n}{2}$, but not when $S$ takes values near the limits of its support. 
    
This observation motivates the use of an approximation to binomial coefficients which combines equations 4.5 and 4.11 in \cite{Odlyzko-1996}; i.e., for all values of $S$ such that $|\frac{n}{2}-S| = o(n^{\frac{2}{3}})$ we have 
\begin{equation*}
    \binom{n}{S} = \frac{2^n}{\sqrt{\frac{1}{2}n\pi}}\exp\bigg\{-\frac{(n-2S)^2}{2n}\bigg\}\exp\{o(1)\},
\end{equation*}
and thus
\begin{equation*}
    \binom{n}{S} \bigg(\frac{1}{2}\bigg)^n = \frac{1}{\sqrt{\frac{1}{2}n\pi}}\exp\bigg\{-\frac{(n-2S)^2}{2n}\bigg\}\exp\{o(1)\}.
\end{equation*} 
The right-hand-side as a continuous function of $S$ exactly equals $\frac{1}{n+1}$ when
\begin{equation*}
    S = \frac{n\pm\sqrt{n\log n}}{2}+o(1)
\end{equation*}
Since $\sqrt{n\log n}=o(n^{\frac{2}{3}})$ (which can be established via L'H\^opital's rule), the approximation is valid for $S$ at and between these critical values.
    
The coefficients $\binom{n}{S}$ decrease for decreasing $S < \frac{n}{2}$ and increasing $S>\frac{n}{2}$, so we have that the desired event occurs when
\begin{equation*}
    \frac{n-\sqrt{n\log n}}{2}+o(1) < S < \frac{n+\sqrt{n\log n}}{2}+o(1),
\end{equation*}
but not when $S$ falls outside this range.
    
Conditioned on $p$, $\frac{S-np}{\sqrt{np(1-p)}}$ converges in law to a standard Gaussian random variable $Z$; hence the desired probability converges to the probability such $Z$ satisfies
\begin{equation*}
    \frac{\sqrt{n}(1-2p)-\sqrt{\log n}}{2\sqrt{p(1-p)}}< Z < \frac{\sqrt{n}(1-2p)+\sqrt{\log n}}{2\sqrt{p(1-p)}}.
\end{equation*}
When $p<(>)\frac{1}{2}$ the bounds are both increasing (decreasing) in order $\sqrt n$, so the interval over which we integrate the standard normal random variable falls increasingly in the upper (lower) tail, and the probability associated with this event vanishes. When $p = \frac{1}{2}$ the $\sqrt{n}$ terms in both bounds vanish, and the lower (upper) bound is decreasing (increasing) in order $\sqrt{\log n}$; the probability that a standard normal falls between these two bounds converges to 1.
\qed

\subsection{Proof of Theorem 3.2}

As in the proof of Theorem 3.1, let $n$ denote the number of possible edges. The Bayes factor selects IE when the number of edges $S = \sum_{i=1}^{n_v}\sum_{j=i}^{n_v} A_{ij}$ satisfies
\begin{equation*}
    \frac{n-\sqrt{n\log n}}{2} \leq S \leq \frac{n+\sqrt{n\log n}}{2}
\end{equation*}
up to some $o(1)$ term which we ignore. The random sum $S$ has a Poisson binomial distribution with success probability vector $\boldsymbol p \in \bigg(\frac{1.5830}{\varepsilon\sqrt n}, 1-\frac{2(1.5830)}{\varepsilon\sqrt n}\bigg)^n$, with
\begin{align*}
    \mathbbm E\lbrack S \rbrack &= \sum_{i=1}^n p_i\\
    \text{var}(S) &= \sum_{i=1}^n p_i(1-p_i).
\end{align*}

Theorem 3.5 of \cite{Tang-2023} provides a uniform upper bound for the Gaussian approximation to the Poisson binomial distribution of $S$, stating that
\begin{equation*}
    \sup_{0\leq k \leq n}\bigg|P(S\leq k) - \Phi\bigg(\frac{k-\sum_{i=1}^n p_i}{\sqrt{\sum_{i=1}^n p_i(1-p_i)}}\bigg{)}\bigg| \leq \frac{0.7915}{\sqrt{\sum_{i=1}^n p_i(1-p_i)}},
\end{equation*}
where $\Phi(z) = P(Z\leq z)$, the standard Gaussian cumulative distribution function. Therefore, the probability that IE is selected when correct satisfies
\begin{align*}
    P\bigg(\frac{n-\sqrt{n\log n}}{2} \leq S \leq \frac{n+\sqrt{n\log n}}{2}\ \big|\  \boldsymbol p\bigg) \geq& \ \Phi\bigg(\frac{n+\sqrt{n\log n}-2\sum_{i=1}^n p_i}{2\sqrt{\sum_{i=1}^n p_i(1-p_i)}}\bigg)\\ &- \Phi\bigg(\frac{n-\sqrt{n\log n}-2\sum_{i=1}^n p_i}{2\sqrt{\sum_{i=1}^n p_i(1-p_i)}}\bigg)\\ &- \frac{1.5830}{\sqrt{\sum_{i=1}^n p_i(1-p_i)}}.
\end{align*}

We next proceed by obtaining a lower bound for the Gaussian probability. The bounds on $\sum_{i=1}^n p_i$ ensure that
\begin{align*}
    \frac{n-\sqrt{n\log n}-2\sum_{i=1}^n p_i}{2\sqrt{\sum_{i=1}^n p_i(1-p_i)}} &<0 < \frac{n+\sqrt{n\log n}-2\sum_{i=1}^n p_i}{2\sqrt{\sum_{i=1}^n p_i(1-p_i)}}.
\end{align*}
Since each $p_i\in (0,1)$, we have $\text{var}(S) = \sum_{i=1}^n p_i(1-p_i) < \frac{n}{4}$; strict inequality is guaranteed, since $\text{var}(S)$ achieves its upper bound only when every $p_i=\frac12$ exactly. We thus have
\begin{align*}
    \frac{n-\sqrt{n\log n}-2\sum_{i=1}^n p_i}{2\sqrt{\sum_{i=1}^n p_i(1-p_i)}} < \frac{n-\sqrt{n\log n}-2n\bar p}{\sqrt{n}} < 0
\end{align*}
and
\begin{align*}
    0< \frac{n+\sqrt{n\log n}-2n\bar p}{\sqrt{n}} < \frac{n-\sqrt{n\log n}-2\sum_{i=1}^n p_i}{2\sqrt{\sum_{i=1}^n p_i(1-p_i)}}.
\end{align*}
The given condition on $\sum p_i$ gives that $\bar p \in \big(\frac{1-\sqrt{n^{-(1+\delta)\log n}}}{2}, \frac{1+\sqrt{n^{-(1+\delta)\log n}}}{2}\big)$, and so we have after some massaging
\begin{equation*}
    \frac{n-\sqrt{n\log n}-2n\bar p}{\sqrt{n}} <  -(1-n^{-\frac{\delta}{2}})\sqrt{\log n} < 0 < (1-n^{-\frac{\delta}{2}})\sqrt{\log n} < \frac{n+\sqrt{n\log n}-2n\bar p}{\sqrt{n}}.
\end{equation*}
Thus the Gaussian probability satisfies
\begin{align*}
    \Phi\bigg(\frac{n+\sqrt{n\log n}-2\sum_{i=1}^n p_i}{2\sqrt{\sum_{i=1}^n p_i(1-p_i)}}\bigg) - \Phi\bigg(\frac{n-\sqrt{n\log n}-2\sum_{i=1}^n p_i}{2\sqrt{\sum_{i=1}^n p_i(1-p_i)}}\bigg)\geq P\big(Z^2 < (1-n^{\frac{-\delta}{2}})^2\log n \big)
\end{align*}

The conditions on the $p_i$, that they all lie in $\bigg(\frac{1.5830}{\varepsilon\sqrt n}, 1-\frac{1.5830}{\varepsilon\sqrt n}\bigg)$, ensures that $- \frac{1.5830}{\sqrt{\sum_{i=1}^n p_i(1-p_i)}} \geq -\varepsilon$. This completes the result.

\subsection{Derivation of the Induced Priors for SBM($\mathbf{X}$)}

we first notice that ER($p$) is a special case of rank-1 block-K SBM($\mathbf{X}$) with latent position $\mathbf{x}=[x_1,\cdots,x_k]^{\mathsf{T}}=[\sqrt{p},\cdots,\sqrt{p}]^{\mathsf{T}}$. As discussed in Section \ref{sec:sbm_known}, it is natural to assign independent beta priors $\text{Beta}(\alpha_i,\beta_i)$ to $x_i$. Letting $\rho_{SBM}$ and $\rho_{ER}$ respectively denote the prior distributions of the rank-1 SBM($\mathbf{X}$) and ER($p$), then the induced prior parameters $(\alpha_i,\beta_i),\ i=1,\cdots,K$ when $p\sim \text{Beta}(\alpha_0,\beta_0)$ should optimize the Kullback-Leibler divergence over the region of the more complex parameter space which corresponds to the simpler parameter space \citep{Klugkist-2007}:
\begin{align*}
    &\min\limits_{\alpha_i,\beta_i} -\mathbbm{E}_{ER}\{{\log {\frac{\rho_{SBM}{([\sqrt{p},\cdots,\sqrt{p}]^{\mathsf{T}})}}{\rho_{ER}(p)}}}\}\\
    =&\min\limits_{\alpha_i,\beta_i} -\mathbbm{E}_{ER}\{\log \frac{\sqrt{p}^{\sum{\alpha_i}-K}(1-\sqrt{p})^{\sum{\beta_i -K}}/\prod^K_{i=1} B(\alpha_i,\beta_i)}{p^{\alpha-1}(1-p)^{\beta-1}/B(\alpha,\beta)}\}\\
    =&\min\limits_{\alpha_i,\beta_i}-\int_0^1 \frac{p^{\alpha-1}(1-p)^{\beta-1}}{B(\alpha,\beta)}\log \frac{\sqrt{p}^{\sum{\alpha_i}-K}(1-\sqrt{p})^{\sum{\beta_i -K}}/\prod^K_{i=1} B(\alpha_i,\beta_i)}{p^{\alpha-1}(1-p)^{\beta-1}/B(\alpha,\beta)}dp
\end{align*}
In the case of a uniform prior for $p$, i.e., $\alpha_0=\beta_0=1$, we can simplify and directly compute the integral; hence the minimization becomes
\begin{align*}
    \min\limits_{\alpha_i,\beta_i} \sum^K_{i=1} \log B(\alpha_i,\beta_i) + \frac{1}{2} \sum^K_{i=1} \alpha_i + \frac{3}{2} \sum^K_{i=1} \beta_i
\end{align*}
which is solved by $\alpha_i=2,\beta_i=1,i=1,\dots,K$.

\subsection{Bernstein-von Mises Theorem on Rank-1 Block-2 SBM} 
In this section we use the Bernstein-von Mises Theorem to show that the posterior of a rank-1 block-2 SBM($[X_1,X_2]$) is approximately Gaussian. Assume that the graph is large enough so that each block has sufficient nodes -- this assumption holds for most of real-world graphs with blocks. The edge existence indicators $a_i,\ i=1,\cdots,n$ fall into three categories: $a_1,\cdots,a_{n_1}$ are distributed as $\text{Bernoulli}(x_1^2)$; $a_{n_1+1},\cdots,a_{n_2}$ are distributed as $\text{Bernoulli}(x_1 x_2)$; $a_{n_2+1},\cdots,a_{n}$ are distributed as $\text{Bernoulli}(x_2^2)$, with $1<n_1<n_2<n$.

Given our parameters $(x_1,x_2)$ with a bivariate beta prior, we first compute the posterior $P_1$ after observing the edges $a_1,\cdots,a_{n_1}$:
$$P_1\propto \text{Beta}(\alpha,\beta)\times x_1^{2\sum_{i=1}^{n_1}a_i}(1-x_1^2)^{\sum_{i=1}^{n_1}(1-a_i)}.$$
To apply the Bernstein-von Mises Theorem to this posterior, we need to check the regularity conditions (\cite{borwanker}):
\begin{enumerate}[(1)]
\item The domain of the parameters is $(0,1)^2$, which is open.
\item The bivariate beta prior is continuous and postive in an open neighborhood of the true parameter.
\item The log-likelihood function for $a_1,\cdots,a_{n_1}$, $ l_1=2\sum_{i=1}^{n_1}a_i\cdot \ln x_1+\sum_{i=1}^{n_1}(1-a_i)\cdot \ln(1-x_1^2)$, can be continuously differentiated twice with respect to $(x_1,x_2)$.
\item Note that $\mathbb{E}[\sum_{i=1}^{n_1}a_i]=n_1 x_1^2$, and $\mathbb{E}[\sum_{i=1}^{n_1}(1-a_i)]=n_1(1-x_1^2)$, so
$$-\mathbb{E}\left(\frac{\partial^2 l_1}{\partial x_1^2}\right)=\mathbb{E}\left(\frac{2\sum_{i=1}^{n_1}a_i}{x_1^2}+\frac{2\sum_{i=1}^{n_1}(1-a_i)(1+x_1^2)}{(1-x_1^2)^2}\right)=\frac{4n_1}{1-x_1^2},$$
which is continuous in $x_1$ and positive and finite.
\end{enumerate}
Therefore, since all the conditions are satisfied, by the theorem, $P_1$ is approximately Gaussian. Now we have an updated belief on the distribution of $(x_1,x_2)$: it becomes $P_1$ from the bivariate beta. Then, given $(x_1,x_2)$ with a $P_1$ prior, we compute the posterior $P_2$ after observing the edges $a_{n_1+1},\cdots,a_{n_2}$:
$$P_2 \propto \text{Beta}(\alpha,\beta)\times x_1^{2\sum_{i=1}^{n_1}a_i}(1-x_1^2)^{\sum_{i=1}^{n_1}(1-a_i)}\times (x_1x_2)^{\sum_{i=n_1+1}^{n_2}a_i}(1-x_1 x_2)^{\sum_{i=n_1+1}^{n_2}(1-a_i)}.$$
To invoke the Bernstein-von Mises Theorem to this posterior $P_2$, we again check the regularity conditions:
\begin{enumerate}[(1)]
\item The domain of the parameters is $(0,1)^2$, which is open.
\item The prior $P_1$ is the product of two continuous non-zero functions, and thus is continuous and positive in an open neighborhood of the true parameter.
\item The log-likelihood function for $a_{n_1+1},\cdots,a_{n_2}$, $l_2=\sum_{i=n_1+1}^{n_2}a_i\cdot \ln(x_1x_2)+\sum_{i=n_1+1}^{n_2}(1-a_i)\cdot\ln(1-x_1x_2)$, can be continuously differentiated twice with respect to $(x_1,x_2)$.
\item Note that $\mathbb{E}[\sum_{i=n_1+1}^{n_2}a_i]=(n_2-n_1)x_1 x_2$, and $\mathbb{E}[\sum_{i=n_1+1}^{n_2}(1-a_i)]=(n_2-n_1)(1-x_1x_2)$,
$$-\mathbb{E}\left(\frac{\partial^2 l_2}{\partial x_1^2}\right)=\mathbb{E}\left(\frac{\sum_{i=n_1+1}^{n_2}a_i}{x_1^2}+\frac{\sum_{i=n_1+1}^{n_2}(1-a_i)x_2^2}{(1-x_1x_2)^2}\right)=\frac{(n_2-n_1)x_2}{x_1(1-x_1x_2)};$$
$$-\mathbb{E}\left(\frac{\partial^2 l_2}{\partial x_2^2}\right)=\mathbb{E}\left(\frac{\sum_{i=n_1+1}^{n_2}a_i}{x_2^2}+\frac{\sum_{i=n_1+1}^{n_2}(1-a_i)x_1^2}{(1-x_1x_2)^2}\right)=\frac{(n_2-n_1)x_1}{x_2(1-x_1x_2)};$$
$$-\mathbb{E}\left(\frac{\partial^2 l_2}{\partial x_1x_2}\right)=\mathbb{E}\left(\frac{\sum_{i=n_1+1}^{n_2}(1-a_i)}{(1-x_1x_2)^2}\right)=\frac{n_2-n_1}{1-x_1x_2};$$
which is continuous in $(x_1,x_2)$ and positive and finite.
\end{enumerate}
Therefore, since all the conditions are satisfied, by the theorem, $P_2$ is approximately Gaussian. Now we have a further updated belief on the distribution of $(x_1,x_2)$: it becomes $P_2$ from $P_1$. Then, given $(x_1,x_2)$ with a $P_2$ prior, we compute the final posterior $p_{x_1,x_2|\mathbf{A}}$ after observing the edges $a_{n_2+1},\cdots,a_{n}$. To invoke the Bernstein-von Mises Theorem to the final posterior $p_{x_1,x_2|\mathbf{A}}$, we again check the regularity conditions:
\begin{enumerate}[(1)]
\item The domain of the parameters is $(0,1)^2$, which is open.
\item The prior $P_2$ is the product of three continuous non-zero functions, and thus is continuous and positive in an open neighborhood of the true parameter.
\item The log-likelihood function for $a_{n_2+1},\cdots,a_{n}$, $ l_3=2\sum_{i=n_2+1}^{n}a_i\cdot \ln x_2+\sum_{i=n_2+1}^{n}(1-a_i)\cdot \ln(1-x_2^2)$, can be continuously differentiated twice with respect to $(x_1,x_2)$.
\item Note that $\mathbb{E}[\sum_{i=n_2+1}^{n}a_i]=(n-n_2) x_2^2$, and $\mathbb{E}[\sum_{i=n_2+1}^{n}(1-a_i)]=(n-n_2)(1-x_2^2)$, so
$$-\mathbb{E}\left(\frac{\partial^2 l_3}{\partial x_2^2}\right)=\mathbb{E}\left(\frac{2\sum_{i=n_2+1}^{n}a_i}{x_2^2}+\frac{2\sum_{i=n_2+1}^{n}(1-a_i)(1+x_2^2)}{(1-x_2^2)^2}\right)=\frac{4(n-n_2)}{1-x_2^2},$$
which is continuous in $x_2$ and positive and finite.
\end{enumerate}
Therefore, all the conditions are satisfied, and by applying the Bernstein-von Mises Theorem three times, we've shown that the posterior of a rank-1 block-2 SBM is approximately Gaussian. Hence, the use of Laplace's method used to approximate this model's evidence is justified.

\subsection{Proof of Theorem 3.3}

For a complete graph, with our choice of prior parameters, i.e., $\text{Beta}(1,1)$ for ER($p$) and $\text{Beta}(1/n,1/n)$ for IE($\mathbf{P}$), we have:
\begin{equation*}
    \log E_{IE} = \sum_{i=1}^n a_i\log\frac{\alpha_i}{\alpha_i+\beta_i} + \sum_{i=1}^n (1-a_i)\log\frac{\beta_i}{\alpha_i+\beta_i}=n\log\frac{1}{2}=\log \frac{1}{2^n},
\end{equation*}
\begin{equation*}
    \log E_{ER} = -\log\bigg[ (n+1) \binom{n}{\sum_{i=1}^na_i}\bigg]= -\log\bigg[ (n+1) \binom{n}{n}\bigg]=\log \frac{1}{n+1}.
\end{equation*}
Note that $\log \frac{1}{2^n}<\log \frac{1}{n+1}$ for any size of graphs, so we only need to compare the evidences of rank-1 block-$K$ SBM($\mathbf{X}$) and ER($p$).

For SBM($\mathbf{X}$), when the graph is complete, we notice that $O_{ij}=0$ for any $i,j$ and $\sum_{i\leq j}S_{ij}=n$. Hence, with prior $\text{Beta}(2,1)$, the evidence of SBM($\mathbf{X}$) is
\begin{align*}
&\int_0^1\cdots\int_0^1 \prod_{i=1}^K \left( x_i^{2S_{ii}+\sum_{j\neq i}S_{ij}} (1-x_i^2)^{O_{ii}} \prod_{i<j}(1-x_ix_j)^{O_{ij}}\right) \prod^K_{i=1}\frac{x_i^{\alpha_i-1}(1-x_i)^{\beta_i-1}}{B(\alpha_i,\beta_i)} dx_1\cdots dx_K\\
&=\int_0^1\cdots\int_0^1 2^K \prod_{i=1}^K \left( x_i^{2S_{ii}+\sum_{j\neq i}S_{ij}+1} \right)  dx_1\cdots dx_K = 2^K\cdot \prod_{i=1}^K \frac{1}{2S_{ii}+\sum_{j\neq i}S_{ij}+2}.
\end{align*}
Note that $\sum^K_{i=1} \left(2S_{ii}+\sum_{j\neq i}S_{ij}+2 \right)= 2n+2K$. Since $2S_{ii}+\sum_{j\neq i}S_{ij}>0$, we use the lemma in Appendix A.8 and get
\begin{equation*}
\prod^K_{i=1} \left(2S_{ii}+\sum_{j\neq i}S_{ij}+2 \right)>2^K\left(\frac{1}{2}(2n+2K)-K+1\right)=2^K(n+1).
\end{equation*}
Hence, when the graph is complete,
\begin{align*}
E_{SBM} = 2^K\cdot \prod_{i=1}^K \frac{1}{2S_{ii}+\sum_{j\neq i}S_{ij}+2}<2^K\cdot \frac{1}{2^K(n+1)}=\frac{1}{n+1}=E_{ER}.
\end{align*}
Therefore, when the graph is complete, the model ER($p$) has the largest evidence among all models in Theorem 3.3, and thus 
is selected in this case.

\subsection{A Lemma used in A.7}
\textbf{Lemma}: For $x_1,x_2,\cdots,x_n$ with $x_i>2,i=1,\cdots,n$, we have $$\prod^n_{i=1}x_i>2^n\left(\frac{1}{2}\sum_{i=1}^n x_i-n+1\right).$$
\textbf{Proof}: We prove this by mathematical induction.

First, when $n=2$, for $x_1>2,x_2>2$, we can write $x_1=2+\epsilon$ and $x_2=x_1+x_2-2-\epsilon$ for some $0<\epsilon<x_1+x_2-4$. Thus,
\begin{align*}
x_1x_2&=(2+\epsilon)(x_1+x_2-2-\epsilon)=2(x_1+x_2)-4-2\epsilon+\epsilon(x_1+x_2)-2\epsilon-\epsilon^2\\
&=2(x_1+x_2)-4+\epsilon(x_1+x_2-4-\epsilon)>2(x_1+x_2)-4 = 2^2\left(\frac{1}{2}(x_1+x_2)-2+1\right).
\end{align*}

Next, assume that the statement in the lemma is true for $n=k$. 

Let's check the case of $n=k+1$. For $x_1,x_2,\cdots,x_{k+1}$ with $x_i>2,i=1,\cdots,k+1$, we have $\sum_{i=1}^{k+1}x_i>\sum_{i=1}^{k}x_i+2$, and $\sum_{i=1}^{k}x_i>2k$. Using the statement for $n=k$, we have:
\begin{align*}
&\prod_{i=1}^{k+1}x_i = \left(\prod_{i=1}^{k}x_i\right)\cdot x_{k+1} = \left(\prod_{i=1}^{k}x_i\right)\left(\sum_{i=1}^{k+1}x_i-\sum_{i=1}^{k}x_i\right)\\
&>2^k\left(\frac{1}{2}\sum_{i=1}^k x_i-k+1\right)\left(\sum_{i=1}^{k+1}x_i-\sum_{i=1}^{k}x_i\right)\\
&=2^{k+1}\left(\frac{1}{2}\sum_{i=1}^k x_i-k+1\right)\left(\frac{1}{2}\sum_{i=1}^{k+1}x_i-\frac{1}{2}\sum_{i=1}^{k}x_i\right)\\
&=2^{k+1}\left(\left(\frac{1}{2}\sum_{i=1}^k x_i-k\right)\frac{1}{2}\sum_{i=1}^{k+1} x_i-\left(\frac{1}{2}\sum_{i=1}^k x_i-k\right)\frac{1}{2}\sum_{i=1}^{k} x_i+\frac{1}{2}\sum_{i=1}^{k+1}x_i-\frac{1}{2}\sum_{i=1}^{k}x_i\right)\\
&>2^{k+1}\left(\left(\frac{1}{2}\sum_{i=1}^k x_i-k\right)\left(\frac{1}{2}\sum_{i=1}^{k} x_i+1\right)-\left(\frac{1}{2}\sum_{i=1}^k x_i-k\right)\frac{1}{2}\sum_{i=1}^{k} x_i+\frac{1}{2}\sum_{i=1}^{k+1}x_i-\frac{1}{2}\sum_{i=1}^{k}x_i\right)\\
&=2^{k+1}\left(\frac{1}{2}\sum_{i=1}^k x_i-k+\frac{1}{2}\sum_{i=1}^{k+1}x_i-\frac{1}{2}\sum_{i=1}^{k}x_i\right) = 2^{k+1}\left(\frac{1}{2}\sum_{i=1}^{k+1}x_i-(k+1)+1\right).
\end{align*}
We thus conclude that the statement holds for any $n>1$, and finish the proof.
\end{document}